\documentclass[twocolumn,twocolappendix]{aastex631}
\usepackage{natbib}
\graphicspath{{fig/}}

\usepackage{amsmath}
\usepackage{algorithm}
\usepackage{amsfonts}
\usepackage{mathrsfs}
\usepackage{amssymb}
\usepackage{color}
\usepackage{graphicx} 
\let\tablenum\relax
\usepackage{siunitx}
\usepackage{comment}

\newcommand{\cha}{\textit{Chandra}}
\newcommand{\XMM}{{XMM-{\it{Newton}}}}

\def \UXClumpy {{\tt \UXClumpy}}
\newcommand{\ms}{\ensuremath{M_{\odot}}}

\begin{document}

\title{Transient Relativistic Iron Emission Line from an X-ray Flaring Supermassive Black Hole}

\author[0000-0002-7791-3671]{Xiurui Zhao}
\affiliation{Cahill Center for Astrophysics, California Institute of Technology, 1216 East California Boulevard, Pasadena, CA 91125, USA}
\affiliation{Department of Astronomy, University of Illinois at Urbana-Champaign, Urbana, IL 61801, USA}

\author[0000-0002-6584-1703]{Marco Ajello}
\affiliation{Department of Physics and Astronomy, Clemson University, Kinard Lab of Physics, Clemson, SC 29634, USA}

\author[0000-0002-2115-1137]{Francesca Civano}
\affiliation{NASA Goddard Space Flight Center, Greenbelt, MD 20771, USA}

\author{Javier A. Garc\'ia}
\affiliation{NASA Goddard Space Flight Center, Greenbelt, MD 20771, USA}
\affiliation{Cahill Center for Astrophysics, California Institute of Technology, 1216 East California Boulevard, Pasadena, CA 91125, USA}

\author[0000-0002-0273-218X]{Elias Kammoun}
\affiliation{Cahill Center for Astrophysics, California Institute of Technology, 1216 East California Boulevard, Pasadena, CA 91125, USA}

\author[0000-0002-2203-7889]{Stefano Marchesi}
\affiliation{Dipartimento di Fisica e Astronomia (DIFA), Università di Bologna, via Gobetti 93/2, I-40129 Bologna, Italy}
\affiliation{Department of Physics and Astronomy, Clemson University, Kinard Lab of Physics, Clemson, SC 29634, USA}
\affiliation{INAF, Osservatorio di Astrofisica e Scienza dello Spazio di Bologna, via P. Gobetti 93/3, 40129 Bologna, Italy}

\author[0000-0003-1659-7035]{Yue Shen}
\affiliation{Department of Astronomy, University of Illinois at Urbana-Champaign, Urbana, IL 61801, USA}

\author[0000-0002-6584-1703]{Daniel Stern}
\affiliation{Jet Propulsion Laboratory, California Institute of Technology, 4800 Oak Grove Drive, Pasadena, CA 91109, USA}

\author[0000-0002-6893-3742]{Qian Yang}
\affiliation{Center for Astrophysics $\vert$ Harvard \& Smithsonian, 60 Garden Street, Cambridge, MA 02138, USA}

\author[0000-0001-9379-4716]{Peter G. Boorman}
\affiliation{Cahill Center for Astrophysics, California Institute of Technology, 1216 East California Boulevard, Pasadena, CA 91125, USA}

\author[0000-0002-4226-8959]{Fiona Harrison}
\affiliation{Cahill Center for Astrophysics, California Institute of Technology, 1216 East California Boulevard, Pasadena, CA 91125, USA}

\author[0000-0003-0172-0854]{Erin Kara}
\affiliation{Department of Physics \& Kavli Institute for Astrophysics and Space Research, Massachusetts Institute of Technology, 70 Vassar St, Cambridge, MA 02139, USA}

\author[0000-0001-6412-2312]{Andrealuna Pizzetti}
\affiliation{European Southern Observatory, Alonso de Córdova 3107, Casilla 19, Santiago, 19001, Chile}

\author[0000-0001-6564-0517]{Ross Silver}
\affiliation{NASA Goddard Space Flight Center, Greenbelt, MD 20771, USA}

\author[0000-0001-5991-6863]{Kirill~V.~Sokolovsky}
\affiliation{Department of Astronomy, University of Illinois at Urbana-Champaign, Urbana, IL 61801, USA}

\author[0000-0002-8501-3518]{Zachary Stone}
\affiliation{Department of Astronomy, University of Illinois at Urbana-Champaign, Urbana, IL 61801, USA}

\author[0000-0003-3638-8943]{Nuria Torres-Alb\`{a}}
\affiliation{Department of Astronomy, University of Virginia, P.O. Box 400325, Charlottesville, VA 22904, USA}

\author[0000-0003-4202-1232]{Qiaoya Wu}
\affiliation{Department of Astronomy, University of Illinois at Urbana-Champaign, Urbana, IL 61801, USA}

\author[0000-0002-6893-3742]{Peixin Zhu}
\affiliation{Center for Astrophysics $\vert$ Harvard \& Smithsonian, 60 Garden Street, Cambridge, MA 02138, USA}

\begin{abstract}
We report the discovery of the first transient relativistic iron K$\alpha$ line in an Active Galactic Nucleus (AGN) J1047+5907. The line was detected 21.5 days (rest-frame) after an X-ray coronal flare observed in 2008 and it exhibits significant broadening consistent with relativistic reflection from the accretion disk in the vicinity of the central supermassive black hole (SMBH). The line has a width of $\sim$300~eV, corresponding to a Keplerian velocity of 14,000 km~s$^{-1}$, at a distance of 5--41 light-days from the SMBH, strongly implying that the observed coronal flare triggered the emergence of the line. This event provides rare direct evidence of the response of the accretion disk to impulsive coronal illumination and offers a new method to probe the SMBH and disk physics. The relativistic modeling favors a broadened line produced by distant reflection from an accretion disk around a rapidly spinning black hole viewed at an intermediate inclination, consistent with other observations. Systematic monitoring of type 1 AGN following strong X-ray flares may open a new observational window into the innermost regions of AGN, enabling constraints on the physics of SMBH and its accretion disk at different radii that are otherwise challenging to access.
\end{abstract}


\section{Introduction} \label{sec:intro}

Active Galactic Nuclei (AGN), powered by accreting matter onto the central supermassive black holes (SMBHs), are among the most luminous persistent astronomical objects in the Universe. The enormous energy released by AGN spans the entire electromagnetic spectrum, from radio to Gamma-rays \citep[e.g.,][]{Elvis1994}. A significant fraction ($\sim$5--20\%) of the AGN luminosity is emitted in X-rays \citep{Lusso2012,Duras2020}, originating from a hot plasma, commonly referred to as the corona, in the immediate vicinity of the inner accretion disk. A fraction of the intrinsic coronal X-ray emission is reflected off the accretion disk, producing fluorescent features, most prominently the iron K$\alpha$ line at $\sim$6.4~keV. When the reflection arises from the inner disk region, the line is skewed and broadened due to relativistic effects induced by the strong gravity near the SMBH. Therefore, the relativistically broadened line can be used as a powerful diagnostic of the physics of the SMBH and the accretion disk in the extreme gravity environment \citep[see][for a review]{Reynolds2003}. Prominent broadened iron K$\alpha$ lines were well detected in only a handful AGN currently \citep[e.g.,][]{Tanaka1995,Fabian2009,Risaliti2013,Jiang2018,Walton2021}, although such features are expected to exist among a large fraction of type 1 (whose optical spectra present broad emission lines) AGN \citep{Nandra2007}.

Although the X-ray variability of AGN has been observed for decades \citep[e.g.,][]{Mushotzky1993,Ulrich1997}, establishing a causal connection between coronal flares and the emergence of relativistic reflection features has been challenging due to observational constraints. Despite extensive studies of relativistic reflection in AGN \citep[see][for a recent review]{Kara2025}, a transient broad iron line directly associated with a discrete coronal flare has not previously been observed. Whether and how the disk promptly responds to impulsive X-ray illumination remains an open question.

Here we present evidence for a transient relativistic iron line in the AGN WISEA J104705.07+590728.4 (hereafter, J1047+5907) at redshift $z$ = 0.391, which emerged about 21.5 days (rest frame) after a coronal flare. The coronal flare might be caused by either magnetic reconnection \citep{Rowan2017}, turbulence \citep{Comisso2019}, or magnetically elevated accretion \citep{Dexter2019}. The line was not detected in observations taken a few years before and after the coronal flare event in October 2008. Its profile is consistent with an origin in disk reflection, representing, to our knowledge, the first reported case of a relativistic line emerging in temporal association with a discrete X-ray flare. This result provides a rare opportunity to investigate the connection between coronal activity and disk reflection and opens a pathway for probing the spatial and temporal structure of the innermost regions of the AGN with future observations.

All uncertainties quoted in this paper are at the confidence level of 1$\sigma$ (68\%) unless otherwise stated. Throughout this work, a flat cosmology is adopted with $H_0$ = 70 km s$^{-1}$ Mpc$^{-1}$, $q_0$ = 0, and $\Lambda$ = 0.73.

\section{Data Reduction and Spectral Analysis}
\subsection{Data Reduction} \label{sec:data_reduction}
The observations were reduced following standard process \citep[e.g., ][]{Zhao2021a} using {\tt CIAO} (V4.15) for \cha\ and {\tt SAS} (V21.0.0) for \XMM. For the \cha\ observations, source spectra were extracted from circular regions with radii of 6\arcsec\ for the first two observations and 23\arcsec\ for the third, corresponding to energy-encircled fractions (EEF) of 90\% at 4~keV at different off-axis angles following the standard choice in \cha/ACIS spectral analysis \cite[e.g.,][]{Civano2016,Evans2024}. Background spectra were extracted from nearby circular regions with radii of 25\arcsec. For the \XMM\ observations, source spectra were extracted using 20\arcsec\ radius regions, corresponding to an EEF of $\sim$75\% at 5~keV following the standard choice in \XMM/EPIC spectral analysis  \cite[e.g.,][]{Barnard2008,Malizia2010,Corral2011}. The background spectra were extracted from nearby 40\arcsec\ circular regions. All spectra were grouped with a minimum of three counts per bin using the HEASoft task \texttt{grppha}.

\subsection{Spectral Analysis}
The spectra were analyzed using \texttt{XSPEC} \citep{Arnaud1996} version 12.13.1. The photoelectric cross-section is from \cite{Verner1996}. The element abundance is from \citet{Anders1989} and metal abundance is fixed to solar. The Cash statistic \citep{Cash1979} was adopted for spectral fitting.

We analyzed the X-ray spectra of the six observations of J1047+5907 using an absorbed power-law model, as the data do not show clear evidence for reflection features from either the torus or the accretion disk. Soft excess emission is apparent in the 2004 September 23 and 2008 November 10 spectra below 1~keV, which may be attributed to the relatively high-quality of these two observations. To account for this, we included a secondary, fractional unabsorbed power-law component, which represents emission that is either not intercepted by the line-of-sight obscurer or is scattered into the line of sight by ionized material. The photon index of this component was tied to that of the primary power law. This scattered component typically contributes less than 10\% of the primary power-law normalization in obscured AGN \citep[e.g.,][]{Marchesi2018}, and we denote its relative normalization as $f_s$. We note that soft X-ray excesses are commonly observed in obscured AGN, although their physical origin remains uncertain. Some proposed explanations include scattered intrinsic coronal emission \citep{Marchesi2018}, thermal emission from the disk \citep{Turner1989}, relativistically blurred reflection \citep{Ballantyne2001}, star formation from the host galaxy \citep{Levenson2001}, or warm compaction \citep{Magdziarz1998}. The limited data quality does not allow for distinguishing between the above scenarios.

In \texttt{XSPEC} nomenclature, the full model is expressed as:
{\tt Model = phabs * (zphabs * zpowerlw + cons * zpowerlw)}, 
where \texttt{zphabs} models the line-of-sight absorption by the obscurer, and \texttt{phabs} accounts for fixed Galactic absorption, set to 8 $\times$ 10$^{19}$ cm$^{-2}$ \citep[\texttt{nh} task,][]{nh}. The best-fit parameters for each epoch are summarized in Table~\ref{Table:best-fit}. Because the photon index is poorly constrained when fit freely across epochs ($\Gamma$ = 1.5$^{+0.3}_{-0.4}$), we fix it at the typical value for obscured AGN of $\Gamma$ = 1.80 \citep{Ricci2017}, for final fittings. Varying the photon index (e.g., $\Gamma = 1.50$--2.10) does not significantly affect the fitted source parameters, particularly the line properties in the 2008 November 10 observation.

\begin{figure*}[tbp] 
\centering
\includegraphics[width=\textwidth]{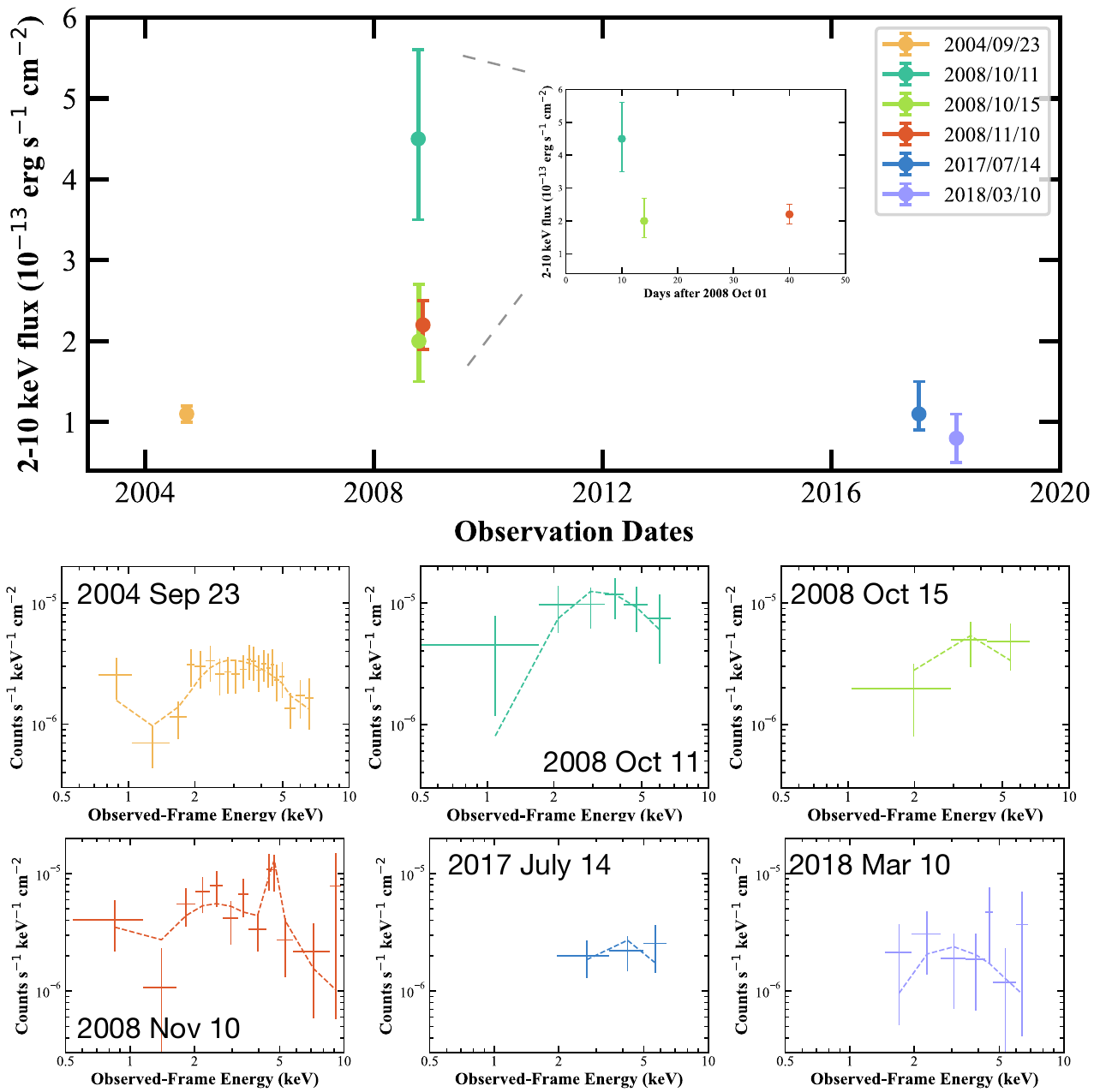}
\vspace{0.2cm}
\caption{Upper: the 2--10~keV X-ray light curve of J1047+5907. The inset is the zoom-in of the flare in 2008. Lower: the X-ray spectra of J1047+5907 from 2004 to 2018. The spectra were  re-grouped so that each bin has as large as 3$\sigma$ detection with maximum three counts in each bin for plotting purpose. The data has been divided by the response effective area.}
\label{fig:light_curve_2_10keV}
\end{figure*}

\begingroup
\renewcommand*{\arraystretch}{1.3}
\begin{table*}
\caption{Information and best-fit results of the six X-ray archival spectra assuming the model described in Section~\ref{sec:fitting}. The listed exposure times correspond to the cleaned exposures after filtering out high-background intervals. Reported net (background-subtracted) counts are measured in the 1--7 keV band for \cha\ and in the 0.5--10 keV band for \XMM\ MOS2. $\Gamma$ is the photon index. N$\rm _{H}$ is the column density in units of $10^{22}$\,cm$^{-2}$. $f_s$ is the scattered fraction in units of 10$^{-2}$. F$_{2-10}$ is the 2--10\,keV flux in units of $10^{-13}$\,erg\,cm$^{-2}$ s$^{-1}$. L$_{2-10}$ is the 2--10\,keV intrinsic luminosity in units of $10^{43}$\,erg\,s$^{-1}$. $^{f}$ is when the parameter is fixed to a given value.}\label{Table:best-fit}
\begin{tabular}{cccccccc}
       \hline
       \hline       
       Date&2004/09/23&2008/10/11&2008/10/15&2008/11/10&2017/07/14&2018/03/10\\
       \hline
       \hline
       Telescope&\cha&XMM&XMM&XMM&\cha&\cha\\
       ObsID&5028&0554120101&0554121001&0554121301&19026&19028\\
       Exp (ks)&71&27&13&37&10&20\\
       Net Counts&191&50&28&121&25&63\\
       \hline
       \hline
       $C$/d.o.f.&38/59&19/15&8/7&37/34&2/6&20/21\\
       $\Gamma$&1.80$^{f}$&1.80$^{f}$&1.80$^{f}$&1.80$^{f}$&1.80$^{f}$&1.80$^{f}$\\
       N$\rm _{H}$&11$_{-2}^{+1}$&11$_{-3}^{+4}$&11$_{-4}^{+7}$&9$_{-2}^{+3}$&19$_{-7}^{+9}$&9$_{-4}^{+13}$\\
	$f_s$&3$_{-1}^{+1}$&0$^{f}$&0$^{f}$&3$_{-1}^{+2}$&0$^{f}$&0$^{f}$\\
       F$_{2-10}$&1.1$_{-0.1}^{+0.1}$&4.5$_{-1.0}^{+1.1}$&2.0$_{-0.5}^{+0.7}$&2.2$_{-0.3}^{+0.3}$&1.1$_{-0.2}^{+0.4}$&0.8$_{-0.3}^{+0.3}$\\
       L$_{2-10}$&8$_{-1}^{+1}$&33$_{-8}^{+11}$&15$_{-5}^{+7}$&15$_{-3}^{+3}$&10$_{-3}^{+5}$&5$_{-2}^{+5}$\\
       \hline
	\hline
\end{tabular}
\end{table*}
\endgroup

\section{Results}\label{sec:results}
J1047+5907 is a radio-quiet, type 1 AGN with multiple broad optical emission lines \citep{Abazajian2009}. The source lies within the Lockman Hole, a region of the sky with a particularly low line-of-sight hydrogen column density, a well-studied area that has been targeted by multiple surveys in X-rays \citep{Wilkes2009} and other wavelengths \citep[e.g.,][]{Mauduit2012}. The source falls within the field-of-view (FoV) of three \cha\ observations and three \XMM\ observations\footnote{In each of the three observations, the source fell only inside the FoV of the \XMM\ MOS2 camera.}, spanning from 2008 to 2018. Detailed information on these six observations is reported in Table~\ref{Table:best-fit}. Archival spectra for each of these six observations were reduced and extracted following standard methods, as detailed in the Supplementary Section~\ref{sec:data_reduction}. 

\subsection{X-ray Spectral Fitting Results}\label{sec:fitting}
We fitted the six spectra using an absorbed power-law model, a common model for characterizing the X-ray spectra of obscured AGN. The model provides a good fit to the spectra from all epochs, with the notable exception of the 2008 November 10 observation, where a prominent broad iron K$\alpha$ line is detected (see Section~\ref{sec:BL} for a detailed analysis). The best-fit model parameters for each observation are summarized in Table~\ref{Table:best-fit}. The spectra and corresponding best-fit models for all epochs are presented in Fig.~\ref{fig:light_curve_2_10keV}.

The best-fit results suggest that J1047+5907 is obscured in X-rays, with a column density of N$_{\rm H}\approx10^{23}$~cm$^{-2}$. This makes J1047+5907 one of the few \citep[$\sim$10\,\%; e.g.,][]{Marchesi2016} type 1 AGN that are obscured in the X-rays with N$_{\rm H}\approx10^{23}$~cm$^{-2}$. X-ray absorption is primarily attributed to gas in the broad-line region (BLR) and to the dusty torus \citep[e.g.,][]{Antonucci1993,Risaliti2007,Markowitz2014,Netzer2015}, while the optical extinction is mainly dominated by the dust within the torus. The observed discrepancy might be explained by a geometrical configuration in which the BLR has a larger covering factor than the torus, allowing the line-of-sight to intersect ionized gas in the BLR without notable dust extinction from the torus. In such a scenario, broad optical emission lines remain detectable, even with persistent X-ray absorption.

The source was in a low-flux state during the 2004, 2017, and 2018 observations, with a 2--10~keV flux of approximately $10^{-13}$\,erg\,cm$^{-2}$ s$^{-1}$. However, J1047+5907 exhibited a significant increase in flux on 2008 October 11 by a factor of $\sim$4 compared with the low-flux level. Four days later, on October 15th, the flux declined rapidly by $\sim$55\%, but was still $\sim$80\% brighter relative to the low-flux level. A subsequent observation on 2008 November 10, approximately one month after the flare (three weeks in rest-frame), exhibited a flux level consistent with the October 15th observation, which was still brighter than the low-flux level. The 2--10~keV light curve of the source is presented in Fig.~\ref{fig:light_curve_2_10keV}.

The intrinsic luminosity of the source in the low-flux state after correcting for absorption is approximately $10^{43}$\,erg\,s$^{-1}$ in the 2--10\,keV rest-frame energy range. During the flare observed in October 2008, its intrinsic luminosity increased and decreased following the same pattern as the observed flux. The X-ray emitting corona is presumed to be a compact structure located within a few to a few tens of gravitational radii ($r_g$ = $GM_{\rm BH}/c^2$) of the SMBH \citep{Chartas2009,Dai2010,Fabian2009,Uttley2014} and is subject to strong general relativistic (GR) effects. The time dilation near the SMBH can shorten the observed flare duration by 10--20\% when the flare originates at radii of 6--10~$r_g$. Such rapid and energetic X-ray transients in AGN could be triggered by magnetic reconnection events in the corona \citep{Rowan2017}, analogous to solar flares but on far larger physical and energetic scales, or by rapid fluctuations of the accretion rate at the inner disk \citep{Stern2018}.

\subsection{The Broad Line in the 2008 November 10 Spectrum}\label{sec:BL}
The \XMM\ spectrum on 2008 November 10 reveals a broad emission feature at rest-frame energies approximately 6--7~keV, which was not detected in other X-ray spectra of J1047+5907. The presence of a broad emission feature near 6.5~keV in the November 8 spectrum suggests the possibility of reprocessed emission from optically thick material, such as an accretion disk, commonly referred to as X-ray reflection. This interpretation is motivated by the fact that the strongest and most ubiquitous features in a reflection spectrum are the inner-shell transitions of iron, observed in the local frame at $\sim$6.4--6.9 keV depending on the ionization state of the gas \citep[e.g.,][]{Garcia2010}. The width of the line further hints at relativistic effects, which can broaden the line profile if the emission originates in the vicinity of the black hole \citep[e.g.,][]{Dauser2010}. A simpler explanation for the broad line width is Doppler broadening due to Keplerian rotation of the disk at a given radius. 

We therefore fit the 2008 November 10 spectrum by adding a Gaussian component to model the broad line feature at 6--7~keV (rest-frame). The spectral shape is now much better modeled and the C-Stat reduced from 37 (with 34 degrees of freedom, d.o.f) to 23 (with 31 d.o.f). The line flux is 4$_{-1}^{+2}$ $\times$ $10^{-14}$\,erg\,cm$^{-2}$ s$^{-1}$, corresponding to a detection significance exceeding $3\,\sigma$. The equivalent width (EW) of the line is 1.9$_{-0.9}^{+1.2}$~keV calculated using the \texttt{eqwidth} tool in \texttt{XSPEC}, indicating a prominent emission line feature. The line is centered at $E_l$ = 6.5$\pm$0.1~keV, consistent with fluorescent Fe K$\alpha$ emission at 6.4~keV. The line has a measured width of $\sigma_l$ = 0.3$\pm0.1$~keV, which is substantially broader than the instrumental resolution ($\sim$0.12~keV) of the \XMM\ MOS2 detector, confirming its broadened nature at greater than 3$\sigma$ significance. The best-fit spectral properties are summarized in Table~\ref{Table:best-fit_Gauss}. It is worth mentioning that the choice of photon index fixing at $\Gamma$ = 1.8 does not significantly alter the properties of the measured emission line (e.g., $E_l$, $\sigma_l$, and line flux F$_l$) as shown in Table~\ref{Table:best-fit_Gauss}, where we also report the best-fit model we obtain by leaving the photon index free to vary for reference. We note that this fit yields a photon index of $\Gamma \sim 3.4$, which is physically implausible for a typical AGN X-ray continuum. This result reinforces our choice to fix $\Gamma$ to a canonical type 1 AGN value.

We performed extensive simulations to assess the likelihood that the detected emission line in the 2008 November 10 spectrum arises from background fluctuations. A total of 10,000 simulated spectra were generated assuming a no-emission-line scenario, and using the best-fit continuum model of the 2008 November 10 observation from Table~\ref{Table:best-fit}. Each simulated spectrum was then fitted with two models: one including only the continuum, and the other adding a Gaussian emission line component with centroid energy $E_{l}$ = 6--7~keV and line width $\sigma_l$ = 0.1--0.5~keV. Among these 10,000 simulations, only one yielded a C-stat improvement exceeding 13.7, which is the $C$-stat improvement when fitting the real data. This corresponds to a false positive probability of $p\sim10^{-4}$, strongly supporting that the detected line is not due to statistical fluctuations.

We applied the same fitting procedure to the remaining five spectra and found that the Gaussian emission line at 6--7~keV is at a significance below 1$\sigma$. Assuming an emission line identical to that observed on 2008 November 10 was present, its flux would be constrained to have an upper limit of $<$1.3 $\times$10$^ {-14}$ erg cm$^{-2}$ s$^{-1}$ (3$\sigma$ confidence) in the highest-quality spectrum, obtained on 2004 September 23. These results indicate that the relativistic line detected on 2008 November 10 is indeed a transient feature.

\subsection{\textit{Swift}-XRT Observations}
The source was observed with \textit{Swift}-XRT twice on 2011 June 14 with a total exposure time of 9~ks, and four additional times between 2014 May 15 and May 25 with a combined exposure of 9~ks. A further \textit{Swift} DDT observation was obtained on 2022 March 9 with an exposure of 2.8~ks. Due to the limited number ($<$10) of source counts in each individual observation, we did not extract spectra but instead converted the net count rates to fluxes, adopting the best-fitting spectral model derived from the 2004 September 23 observation. The inferred 2--10 keV fluxes are 1.3$^{+0.8}_{-0.5}$, 1.8$^{+0.9}_{-0.6}$, and 1.4$^{+1.9}_{-0.9}$ $\times$ 10$^{-13}$~erg~cm$^{-2}$~s$^{-1}$ for the 2011, 2014, and 2022 observations, respectively. These values are consistent with the source’s low-flux state within uncertainties.

\begingroup
\renewcommand*{\arraystretch}{1.3}
\begin{table}
\centering
\caption{Best-fit results of the \XMM\ 2008/11/10 observation after adding the broad emission line component assuming frozen and free photon indices. {\tt E$_f$} is the centroid energy of the line and {\tt $\sigma_l$} is the standard deviation of the Gaussian model. {\tt F$_f$} is the line flux at rest-frame 5.5--7.5~keV. {\tt EW$_l$} is the equivalent width of the line. $^u$ indicates that the upper limit is unconstrained.}
\label{Table:best-fit_Gauss}
  \begin{tabular}{llcc}
       \hline
       \hline       
       Model&Unit&Gaussian$_{freeze\,\Gamma}$&Gaussian$_{free\,\Gamma}$\\
       \hline
       \hline
       $C$/d.o.f.&&23.4/31&21.1/30\\
       $\Gamma$&&1.80$^{f}$&3.4$^{+1.3}_{-1.1}$\\
       N$\rm _{H}$&$10^{22}$\,cm$^{-2}$&7$\pm$2&11$^{+4}_{-3}$\\
	$f_s$&10$^{-2}$&5$_{-2}^{+3}$&0.5$_{-0.4}^{+1.5}$\\
	$E_{l}$&keV&6.5$\pm$0.1&6.5$\pm$0.1\\
	$\sigma_{l}$&keV&0.3$\pm$0.1&0.3$\pm$0.1\\
	F$_{l}$&$10^{-14}$\,cgs&4$_{-1}^{+2}$&5$_{-1}^{+2}$\\
	EW$_{l}$&keV&1.9$_{-0.9}^{+1.2}$&3.4$_{-0.3}^{+u}$\\
       \hline
	\hline
\end{tabular}
\end{table}
\endgroup

\begin{figure*}[htbp] 
\centering
\includegraphics[width=\textwidth]{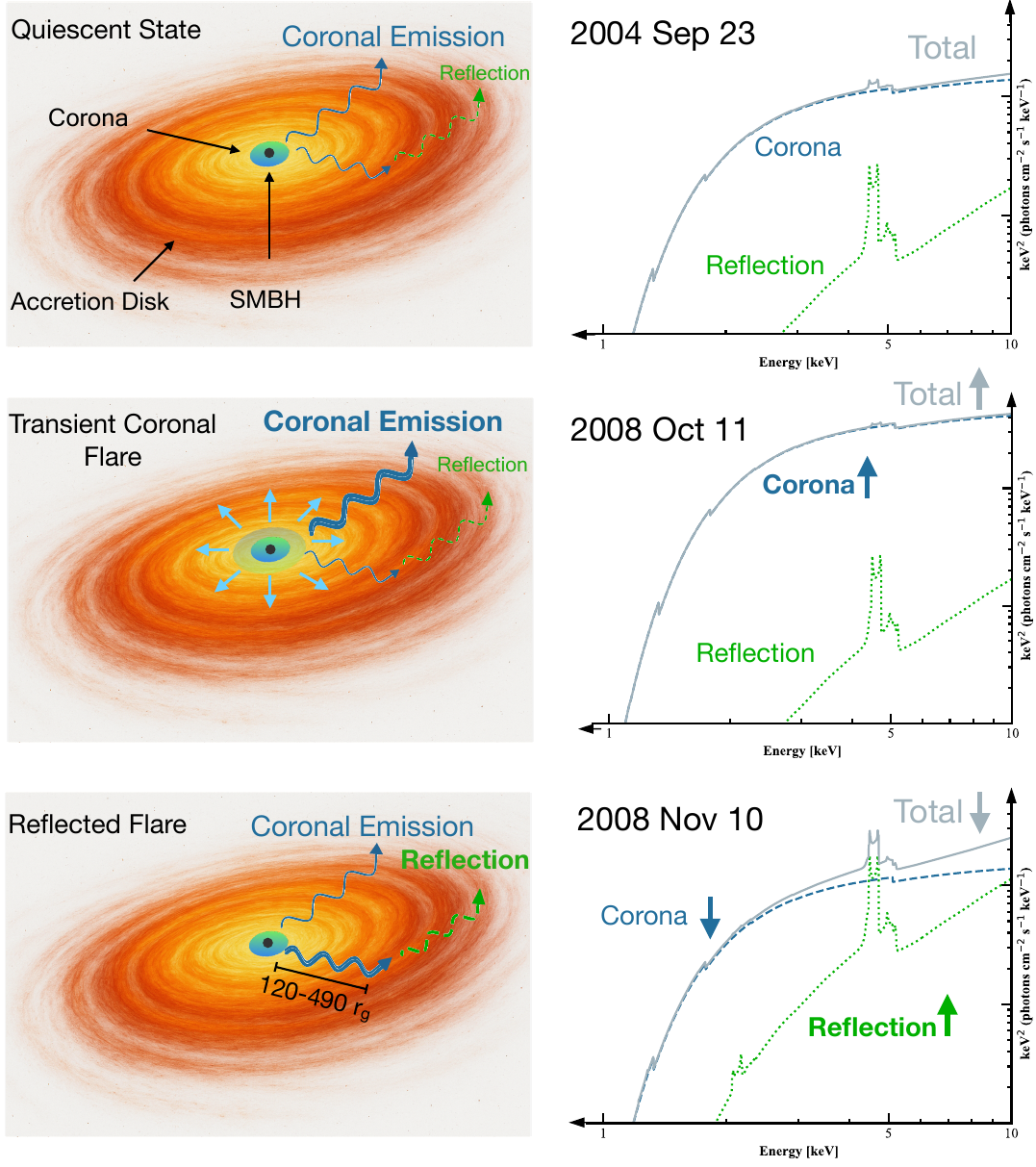}
\vspace{0.2cm}
\caption{Schematic illustration of the relativistic iron line produced by a coronal flare. The top panel represents the source in a low-flux state, as observed on 2004 September 23. The middle panel shows the onset of a transient coronal flare, during which both the coronal emission and total X-ray flux increase significantly. The bottom panel depicts the subsequent phase, when the flare has propagated to distances of several hundred gravitational radii and is reflected by the accretion disk, leading to a substantial enhancement in the reflected component and the emergence of a relativistically broadened iron K$\alpha$ line.}
\label{fig:illustration}
\end{figure*}

\section{Discussion}\label{sec:discussion}
Broad iron K$\alpha$ emission lines are typically interpreted as the result of coronal X-ray emission being reflected off high-velocity gas in the inner accretion disk, where relativistic effects near the SMBH shape the line profile \citep{Fabian2000}. In the case of J1047+5907, such a broad line is detected only in the 2008 November 10 observation (about one month after the flare) and is absent in all other epochs. This transient behavior suggests that the line was triggered by a strong coronal flare that occurred prior to the 2008 November observation, which is temporally consistent with the flare detected. As the flare irradiated the accretion disk, it was reflected and subsequently observed as the relativistically broadened line.

In the low-flux state, the reflected disk emission remains undetectable due to its much lower luminosity compared to the direct coronal continuum along the line of sight (illustrated in Fig.~\ref{fig:illustration}, top panel). During the flare on 2008 October 11, the coronal flux increased by a factor of approximately 4; however, the reflected component was likely still overwhelmed by the intense coronal continuum, and the extreme relativistic effects significantly smear the line profile of the reflected component (Fig.\ref{fig:illustration}, middle panel). 
By 2008 November 10, approximately one month after the observed flare onset, the intrinsic coronal emission had returned to its low-flux level, while the flare itself had propagated outward and illuminated more distant ($\sim$21.5~light-days considering the redshift) regions of the disk. At these larger radii, the reflected emission was stronger in contrast, allowing the iron line to emerge prominently above the continuum (Fig.~\ref{fig:illustration}, bottom panel).

To estimate the time at which the coronal flare occurred, we calculate the distance from the central SMBH (assuming that the corona is close to the SMBH) to the region of the disk responsible for reflecting the flare emission ($D$), inferred from the observed width of the relativistic iron line. The line broadening is primarily due to the Doppler motion of high-velocity gas in the disk. Assuming a geometrically thin disk with gas in Keplerian rotation, the orbital velocity at a radius $D$ = $n\,R_g$ is approximately $v$ = $c/\sqrt{n}$, where $c$ is the speed of light. 
The observed velocity ($v_o$) depends on the inclination angle $\theta$ between the disk normal and the line-of-sight, and is given by $v_o$ = $v\times$sin$\theta$/$\sqrt{2}$. For a face-on view, $v_o\approx$\,0, while for an edge-on view, $v_o\approx\,v/\sqrt{2}$. The observed velocity is related to the spectral line properties through $v_o/c$ = $\sigma_l/E_l$. 

Using the observed values $E_l$ = 6.5~keV and $\sigma_l$ = 0.3$\pm$0.1~keV, we find $v_o\approx$0.045$\pm$0.015~$c$. This implies $v\approx$0.064$\pm$0.021/sin$\theta$~$c$. Since the line-of-sight intersects the BLR, the inclination angle must exceed the BLR opening angle. For type 1 AGN, the BLR covering factor corresponds to cos$\theta_{BLR}\sim$0.1--0.3 or $\theta_{BLR}$ = 73$^\circ$--84$^\circ$ \citep{Maiolino2001,Pandey2023}, implying $\theta\gtrsim$73$^\circ$ and sin$\theta\gtrsim$0.96, consistent with the intermediate inclination rather than the typical face-on inclination in type 1 AGN. Substituting this into the expression for $n$ yields an estimated disk reflection radius of $D\sim$220$^{+270}_{-100}$ $r_g$. Given the large line width and the inferred emission radius, it is unlikely that the line originates from the canonical BLR, which is typically located at radii of $\sim$10$^4$--10$^5$ $r_g$ for AGN with luminosity like J1047+5907 \citep[e.g.,][]{Kaspi2005,Shen2024}.

The light travel time from the SMBH to the reflecting region of the disk can be estimated as $t$ = $D/c$ = $n\times GM_{\rm BH}/c^3$. For J1047+5907, the black hole mass is estimated to be M$_{\rm BH}\sim$10$^9$~M$_{\odot}$ based on the single-epoch optical spectroscopy \citep{Shen2011}. This method carries systematic uncertainties of a factor of $\sim$2.5 \citep{Shen2013}, implying a plausible black hole mass of J1047+5907 being M$_{\rm BH}\sim$0.4--2.5$\times$10$^9$~M$_{\odot}$. The light-crossing time for 1~$r_g$ corresponding to a black hole mass of 10$^9$~\ms\ is about 1.4~hours. Substituting this into the light-travel time calculation, we find that $t$ lies in the range of 13.3$_{-8.9}^{+31.5}$ days in the rest frame, corresponding to approximately 18.5$_{-12.5}^{+43.5}$ days in the observed frame. 

Under the scenario described above, the emission line observed on 2008 November 10 must have originated from a coronal flare that occurred between 2008 September 9 and November 4, based on the estimated light-travel time. The flare detected around 2008 October 10 is therefore a strong candidate for the event that illuminated the accretion disk and produced the observed broad line profile, although we cannot rule out the possibility of an additional flare occurring between these dates.

It is worth noting that a purely Doppler-broadened origin would produce a classical, double-peaked, and symmetric line profile. However, such a scenario fails to reproduce the data, as the observed feature is distinctly asymmetric (Fig.~\ref{fig:relxill}), with enhanced emission on the blue side (i.e. at higher energies). The skewed profile instead points to the influence of relativistic effects and disk reflection. More physically motivated models such as {\sc relxill} \citep{Garcia2014}, which self-consistently incorporate Doppler, special- and general-relativistic effects as well as the geometry and physics of the accretion disk, are therefore required. 

Direct fits with {\sc relxill} \citep{Garcia2014} tend to overfit the spectra, due to the limited data quality of the 2008 November 10 observation, resulting in poorly constrained parameters. We therefore restrict ourselves to exploring the conditions under which {\sc relxill} can reproduce a line profile broadly consistent with the observations. Comparison of {\sc relxill} models across a range of parameter combinations with the November 8 data indicates a strong reflection signal, produced by the reprocessing of X-rays in optically thick material such as an accretion disk. The emission profile is consistent with inner-shell transitions of iron, and the line shape suggests reflection arising at radii $>$21.5~$R_{\rm g}$ (at the 3$\sigma$ confidence level) from the black hole. The reflection component dominates the spectrum, as the data can be reproduced without requiring an additional power-law continuum. Fits with relativistic reflection models generally favor high black hole spin, reflection from distant material, and intermediate inclinations ($\sim50^{^\circ}$, where face-on is $0^{^\circ}$). In Fig.~\ref{fig:relxill} of Section~\ref{sec:relxill}, we illustrate how the line profile varies with the ionization parameter, disk inclination, and inner disk radius, as discussed above.

\section{Conclusion}\label{sec:conclusion}
We report the discovery of the first transient relativistic X-ray iron K$\alpha$ line in an AGN, along with the detection of the preceding coronal flare that likely triggered the reflection event. While the limited data quality prevents tight constraints on the physical parameters of the SMBH and accretion disk in J1047+5907, this event highlights the potential of time-domain X-ray spectroscopy to probe SMBH spin and accretion disk structures. Our results suggest that cadenced monitoring of AGN exhibiting transient X-ray flares could become a powerful tool for studying the disk properties, particularly plasma densities and inclinations at different radii. Such observations can be enabled by current facilities, including joint campaigns with {\it Einstein Probe} \citep{Yuan2025} and \XMM, and will be significantly enhanced by upcoming missions such as {\it eXTP} \citep{Zhang2025} and {\it NewAthena} \citep{Cruise2025}, and the proposed {\it AXIS} concept \citep{Reynolds2023}.

\section{Acknowledgements}
The authors thank the anonymous referee for their insightful comments on the manuscript. XZ acknowledges the support from NASA funding 80NSSC24K1031. We appreciate the Targets of Opportunity observation from the Neil Gehrels \textit{Swift} Observatory. This research has made use of data obtained from the \cha\ Data Archive and software provided by the \cha\ X-ray Center (CXC). This work is based on observations obtained with \XMM, an ESA science mission with instruments and contributions directly funded by ESA Member States and NASA. This work makes use of the data from SDSS. Funding for the Sloan Digital Sky Survey has been provided by the Alfred P. Sloan Foundation, the U.S. Department of Energy Office of Science, and the Participating Institutions. 

This paper employs a list of Chandra datasets, obtained by the Chandra X-ray Observatory, contained in\dataset[DOI: 10.25574/cdc.517]{https://doi.org/10.25574/cdc.517}
\facilities{\cha, \XMM}

\appendix
\renewcommand{\thesection}{APPENDIX~\Alph{section}}

\begingroup
\renewcommand*{\arraystretch}{1.3}
\begin{table*}
\centering
\caption{Best-fit results of the 2008 November 10 observation assuming three iron emission line components. $^{f}$ is when the parameter is fixed to a given value, and $^{u}$ is when the parameter is unconstrained.}
\label{Table:Tripple_Gauss}
  \begin{tabular}{llcc}
       \hline
       \hline       
       Model&Unit&Triple Gaussians&Triple Narrow Gaussians\\
       \hline
       \hline
       $C$/d.o.f.&&21.5/29&24.5/31\\
       $\Gamma$&&1.80$^{f}$&1.80$^{f}$\\
       N$\rm _{H}$&$10^{22}$\,cm$^{-2}$&7$\pm2$&8$\pm2$\\
	$f_s$&10$^{-2}$&5$_{-2}^{+3}$&4$_{-2}^{+3}$\\
	\hline
	E$_{\rm Fe\,K\alpha}$&keV&6.4$^{f}$&6.4$^{f}$\\
	$\sigma_{\rm Fe\,K\alpha}$&keV&0.3$\pm0.1$&0.01$^{f}$\\
	F$_{\rm Fe\, K\alpha}$&$10^{-14}$\,cgs&3$_{-2}^{+1}$&1.1$_{-0.7}^{+0.9}$\\
	EW$_{\rm Fe\, K\alpha}$&keV&0.6$_{-0.4}^{+0.9}$&0.15$_{-0.10}^{+0.16}$\\
	\hline
	E$_{\rm Fe\, XXV}$&keV&6.7$^{f}$&6.7$^{f}$\\
	$\sigma_{\rm Fe\, XXV}$&keV&0.1$_{-u}^{+0.1}$&0.01$^{f}$\\
	F$_{\rm Fe\, XXV}$&$10^{-14}$\,cgs&2$\pm1$&2$\pm1$\\
	EW$_{\rm Fe\, XXV}$&keV&0.4$_{-0.4}^{+0.6}$&0.4$_{-0.2}^{+0.3}$\\
	\hline
	E$_{\rm Fe\, XXVI}$&keV&6.97$^{f}$&6.97$^{f}$\\
	$\sigma_{\rm Fe\, XXVI}$&keV&0.01$^{f}$&0.01$^{f}$\\
	F$_{\rm Fe\, XXVI}$&$10^{-14}$\,cgs&0$_{-u}^{+0.8}$&0.2$_{-u}^{+0.9}$\\
	EW$_{\rm Fe\, XXVI}$&keV&0$_{-u}^{+0.4}$&0$_{-u}^{+0.15}$\\ 
       \hline
	\hline
\end{tabular}
\end{table*}
\endgroup

\section{A Single Line or Multiple Iron Lines}
The most prominent emission feature near 6.5 keV is consistent with the Fe K$\alpha$ line at 6.4 keV, typically associated with reflection from neutral material. The observed blueshift of the line centroid suggests either a modest degree of ionization in the emitting gas or relativistic distortion of the line profile. Other potential contributors near this energy include the Fe XXV and Fe XXVI lines at 6.7 keV and 6.97 keV, respectively, although these are generally much weaker than the neutral Fe K$\alpha$ feature. 

To investigate possible contributions from these ionized lines, we performed an additional fit including two Gaussian components at the expected energies of Fe XXV and Fe XXVI. The results indicate a tentative detection of the Fe XXV line, though its width remains unconstrained, while no significant Fe XXVI emission is detected (we therefore fix its line-width at 0.01~keV to better constrain its line flux and equivalent width). Importantly, the Fe K$\alpha$ line remains robustly detected and exhibits a similar width to that obtained from the single-Gaussian model, suggesting that the observed broad feature indeed includes a substantial component from either neutral or mildly ionized iron.

The improvement in the fit from adding the two additional Gaussian components is negligible, with a $\Delta$C-stat of 1.9. The best-fit values for the triple Gaussian model are reported in Table~\ref{Table:Tripple_Gauss}. Due to the limited statistical improvement, we adopt the single-Gaussian model in the main analysis. Adopting the Fe K$\alpha$ line parameters from the triple-Gaussian model does not alter the overall results or conclusions of the analysis.

Another scenario for the observed broad line profile is that the line is composed of a blend of narrow, unresolved iron lines, produced by the distant, partially ionized gas in the broad line region (BLR) or the torus induced by the AGN flare (as the peak of the X-ray flare in 2008 is uncertain). We found that the observed line profile cannot be adequately described by a combination of narrow, unresolved ($\sim$0.01 keV) emission lines from Fe K$\alpha$, Fe XXV, and Fe XXVI (Table~\ref{Table:Tripple_Gauss}). Instead, models that include a broad Fe K$\alpha$ component provide a better fit to the observed spectrum, indicating that a broad Fe K$\alpha$ line is preferred. In addition, the line flux of Fe K$\alpha$ is found to be fainter than the line flux of Fe XXV under this scenario. However, the Fe K$\alpha$ line flux is typically much larger than the Fe XXV line flux in different known obscured AGN with much better spectral quality \citep[e.g.,][]{Pounds2006}, although we cannot exclude the possibility that the reflector is highly ionized. In addition, the narrow line scenario suggests a distant reflector that is $\gtrsim$90 light-days (rest-frame) away from the SMBH (assuming an unresolved line width of $\sigma_l$ = 0.05$\pm$0.05~keV). The flare around 2008 October 11 and October 15 suggests that the flare evolve very fast. Therefore, if the observed line profile is produced by a distant reflector, the peak luminosity should be much higher (a few tens or a few hundreds times brighter) than the source luminosity on 2008 October 11. Such significant variability has not been observed in the previous observations of AGN with a black hole mass of M$_{\rm BH}\sim$10$^9$~M$_{\odot}$. Therefore, the reflection from the disk rather than the distant material is favored and discussed in the main text.

In addition, we tested whether the detected feature could be described as a blend of mildly ionized iron emission lines, with rest-frame energies in the range E$_{\ell}$ = 6.4--6.7~keV. We fitted the spectrum with an absorbed power-law continuum plus two narrow Gaussian components ($\sigma_{\ell}$ = 0.01~keV), allowing the line centroids to vary freely. The best-fit model ($C$/{d.o.f.} = 20.9/30) yields centroid energies of $\sim$6.30~keV and $\sim$6.76~keV. This configuration does not support an interpretation as a blend of mildly ionized iron lines; instead, the skewed line profile is favored by relativistic effects arising from strong gravity in the vicinity of the SMBH, as discussed in \ref{sec:relxill}.

\begin{figure*}[htbp] 
\centering
\includegraphics[width=\textwidth]{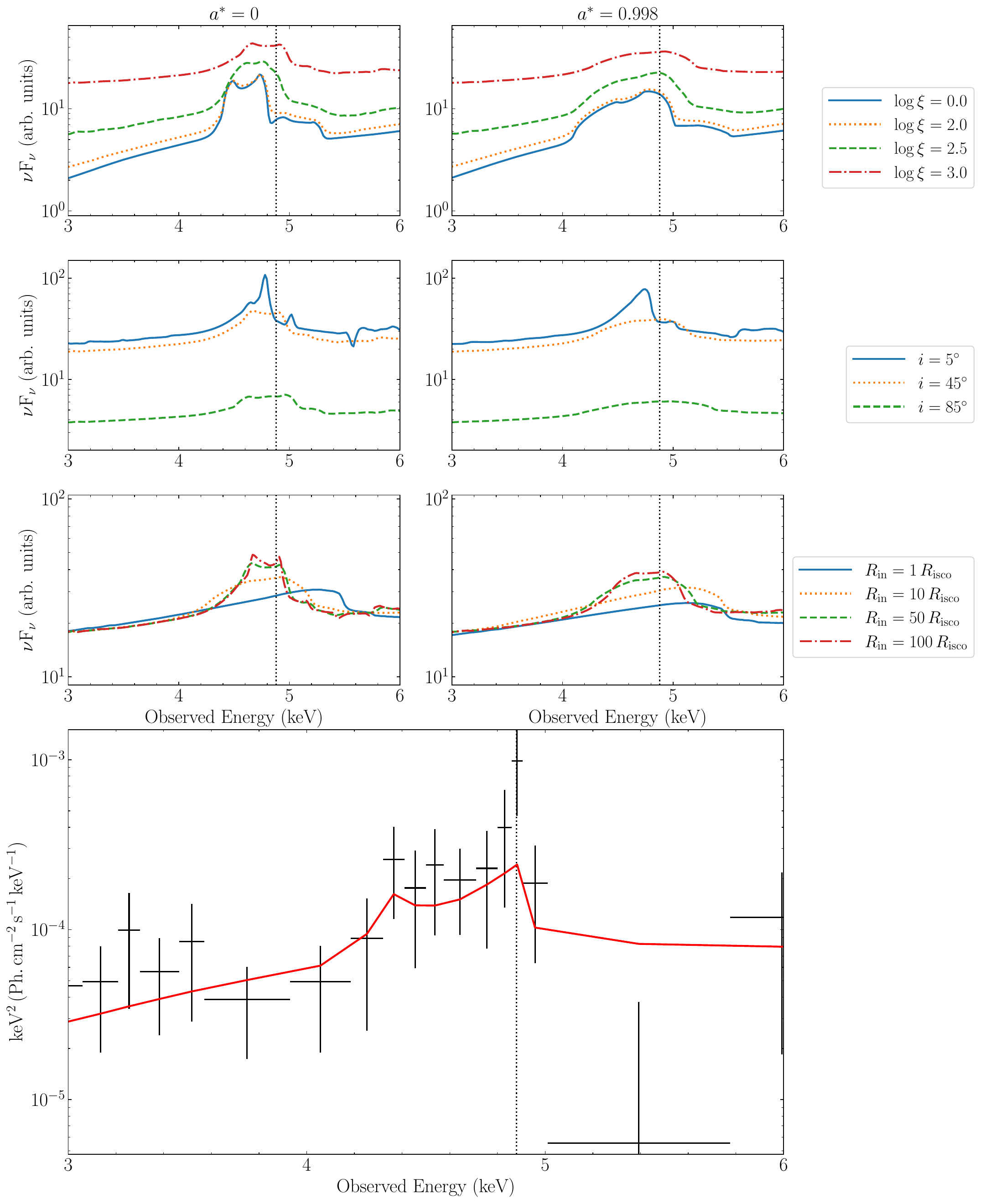}
\vspace{0.2cm}
\caption{Upper: Effect of key reflection parameters, ionization, disk inclination, and inner disk radius (top to bottom), on the Fe K$\alpha$ line profile. Left and right panels show the profiles for a non-spinning and a maximally positively spinning black hole, respectively.
Lower: Zoom in spectrum from the 2008 November 10 observation with one of the realization of the best-fit from {\sc relxill}.}
\label{fig:relxill}
\end{figure*}   

\section{Analyzing the Broad Line with the Relativistic Model} \label{sec:relxill}
The left and right panels show profiles for a non-rotating and a maximally rotating black hole, respectively. The bottom panel presents the observed spectrum fitted with a reflection model, where we fix the inclination to $50^{\circ}$, the inner disk radius to $200,R_{\rm g}$, and adopt a low ionization parameter. We stress that higher-quality observations will be helpful for placing tighter limits on the physical parameters. 

To confirm that relativistic broadening is required, we modeled the spectrum using {\tt xillver} \citep{Garcia2013}. Like {\sc relxill}, {\tt xillver} accounts for the disk reflection spectrum, but {\tt xillver} did not include the relativistic effects. The {\tt xillver} model failed to characterize the observed asymmetric, broadened emission profile, yielding significant residuals in the Fe K$\alpha$ band. This discrepancy strongly suggests that relativistic blurring is essential to explain the observed line properties.

{\bf It has been shown that the high accretion-disk plasma density can influence the resulting spectral shape \citep{Ding_2024}. However, high-density effects have so far only been explored in {\tt xillver}, not in {\sc relxill}. Moreover, variations in disk density produce only minor changes in the Fe K$\alpha$ line profile \citep[Fig.~3 in][]{Ding_2024}. We therefore do not further consider disk-density effects in deriving the black hole and disk parameters.}

\bibliography{referencezxr}{}

@article{Ding_2024,
	author = {Ding, Yuanze and Garcıa, Javier A. and Kallman, Timothy R. and Mendoza, Claudio and Bautista, Manuel and Harrison, Fiona A. and Tomsick, John A. and Dong, Jameson},
	doi = {10.3847/1538-4357/ad76a1},
	journal = {The Astrophysical Journal},
	month = {oct},
	number = {2},
	pages = {280},
	publisher = {The American Astronomical Society},
	title = {Next-generation Accretion Disk Reflection Model: High-density Plasma Effects},
	url = {https://doi.org/10.3847/1538-4357/ad76a1},
	volume = {974},
	year = {2024}}

@article{Shen2024,
	adsurl = {https://ui.adsabs.harvard.edu/abs/2024ApJS..272...26S},
	archiveprefix = {arXiv},
	author = {{Shen}, Yue and {Grier}, Catherine J. and {Horne}, Keith and {Stone}, Zachary and {Li}, Jennifer I. and {Yang}, Qian and {Homayouni}, Yasaman and {Trump}, Jonathan R. and {Anderson}, Scott F. and {Brandt}, W.~N. and {Hall}, Patrick B. and {Ho}, Luis C. and {Jiang}, Linhua and {Petitjean}, Patrick and {Schneider}, Donald P. and {Tao}, Charling and {Donnan}, Fergus. R. and {AlSayyad}, Yusra and {Bershady}, Matthew A. and {Blanton}, Michael R. and {Bizyaev}, Dmitry and {Bundy}, Kevin and {Chen}, Yuguang and {Davis}, Megan C. and {Dawson}, Kyle and {Fan}, Xiaohui and {Greene}, Jenny E. and {Gr{\"o}ller}, Hannes and {Guo}, Yucheng and {Ibarra-Medel}, H{\'e}ctor and {Jiang}, Yuanzhe and {Keenan}, Ryan P. and {Kollmeier}, Juna A. and {Lejoly}, Cassandra and {Li}, Zefeng and {de la Macorra}, Axel and {Moe}, Maxwell and {Nie}, Jundan and {Rossi}, Graziano and {Smith}, Paul S. and {Tee}, Wei Leong and {Weijmans}, Anne-Marie and {Xu}, Jiachuan and {Yue}, Minghao and {Zhou}, Xu and {Zhou}, Zhimin and {Zou}, Hu},
	doi = {10.3847/1538-4365/ad3936},
	eid = {26},
	eprint = {2305.01014},
	journal = {\apjs},
	month = jun,
	number = {2},
	pages = {26},
	primaryclass = {astro-ph.GA},
	title = {{The Sloan Digital Sky Survey Reverberation Mapping Project: Key Results}},
	volume = {272},
	year = 2024}

@article{Kaspi2005,
	adsurl = {https://ui.adsabs.harvard.edu/abs/2005ApJ...629...61K},
	archiveprefix = {arXiv},
	author = {{Kaspi}, Shai and {Maoz}, Dan and {Netzer}, Hagai and {Peterson}, Bradley M. and {Vestergaard}, Marianne and {Jannuzi}, Buell T.},
	doi = {10.1086/431275},
	eprint = {astro-ph/0504484},
	journal = {\apj},
	month = aug,
	number = {1},
	pages = {61-71},
	primaryclass = {astro-ph},
	title = {{The Relationship between Luminosity and Broad-Line Region Size in Active Galactic Nuclei}},
	volume = {629},
	year = 2005}

@article{Garcia2013,
	adsurl = {https://ui.adsabs.harvard.edu/abs/2013ApJ...768..146G},
	archiveprefix = {arXiv},
	author = {{Garc{\'\i}a}, J. and {Dauser}, T. and {Reynolds}, C.~S. and {Kallman}, T.~R. and {McClintock}, J.~E. and {Wilms}, J. and {Eikmann}, W.},
	doi = {10.1088/0004-637X/768/2/146},
	eid = {146},
	eprint = {1303.2112},
	journal = {\apj},
	month = may,
	number = {2},
	pages = {146},
	primaryclass = {astro-ph.HE},
	title = {{X-Ray Reflected Spectra from Accretion Disk Models. III. A Complete Grid of Ionized Reflection Calculations}},
	volume = {768},
	year = 2013}

@article{Antonucci1993,
	adsurl = {https://ui.adsabs.harvard.edu/abs/1993ARA&A..31..473A},
	author = {{Antonucci}, Robert},
	doi = {10.1146/annurev.aa.31.090193.002353},
	journal = {\araa},
	month = jan,
	pages = {473-521},
	title = {{Unified models for active galactic nuclei and quasars.}},
	volume = {31},
	year = 1993}

@article{Markowitz2014,
	adsurl = {https://ui.adsabs.harvard.edu/abs/2014MNRAS.439.1403M},
	archiveprefix = {arXiv},
	author = {{Markowitz}, A.~G. and {Krumpe}, M. and {Nikutta}, R.},
	doi = {10.1093/mnras/stt2492},
	eprint = {1402.2779},
	journal = {\mnras},
	month = apr,
	number = {2},
	pages = {1403-1458},
	primaryclass = {astro-ph.GA},
	title = {{First X-ray-based statistical tests for clumpy-torus models: eclipse events from 230 years of monitoring of Seyfert AGN}},
	volume = {439},
	year = 2014}

@article{Netzer2015,
	adsurl = {https://ui.adsabs.harvard.edu/abs/2015ARA&A..53..365N},
	archiveprefix = {arXiv},
	author = {{Netzer}, Hagai},
	doi = {10.1146/annurev-astro-082214-122302},
	eprint = {1505.00811},
	journal = {\araa},
	month = aug,
	pages = {365-408},
	primaryclass = {astro-ph.GA},
	title = {{Revisiting the Unified Model of Active Galactic Nuclei}},
	volume = {53},
	year = 2015}

@article{Risaliti2007,
	adsurl = {https://ui.adsabs.harvard.edu/abs/2007ApJ...659L.111R},
	archiveprefix = {arXiv},
	author = {{Risaliti}, G. and {Elvis}, M. and {Fabbiano}, G. and {Baldi}, A. and {Zezas}, A. and {Salvati}, M.},
	doi = {10.1086/517884},
	eprint = {astro-ph/0703173},
	journal = {\apjl},
	month = apr,
	number = {2},
	pages = {L111-L114},
	primaryclass = {astro-ph},
	title = {{Occultation Measurement of the Size of the X-Ray-emitting Region in the Active Galactic Nucleus of NGC 1365}},
	volume = {659},
	year = 2007}

@article{Abazajian2009,
	adsurl = {https://ui.adsabs.harvard.edu/abs/2009ApJS..182..543A},
	archiveprefix = {arXiv},
	author = {{Abazajian}, Kevork N. and {Adelman-McCarthy}, Jennifer K. and {Ag{\"u}eros}, Marcel A. and {Allam}, Sahar S. and {Allende Prieto}, Carlos and {An}, Deokkeun and {Anderson}, Kurt S.~J. and {Anderson}, Scott F. and {Annis}, James and {Bahcall}, Neta A. and {Bailer-Jones}, C.~A.~L. and {Barentine}, J.~C. and {Bassett}, Bruce A. and {Becker}, Andrew C. and {Beers}, Timothy C. and {Bell}, Eric F. and {Belokurov}, Vasily and {Berlind}, Andreas A. and {Berman}, Eileen F. and {Bernardi}, Mariangela and {Bickerton}, Steven J. and {Bizyaev}, Dmitry and {Blakeslee}, John P. and {Blanton}, Michael R. and {Bochanski}, John J. and {Boroski}, William N. and {Brewington}, Howard J. and {Brinchmann}, Jarle and {Brinkmann}, J. and {Brunner}, Robert J. and {Budav{\'a}ri}, Tam{\'a}s and {Carey}, Larry N. and {Carliles}, Samuel and {Carr}, Michael A. and {Castander}, Francisco J. and {Cinabro}, David and {Connolly}, A.~J. and {Csabai}, Istv{\'a}n and {Cunha}, Carlos E. and {Czarapata}, Paul C. and {Davenport}, James R.~A. and {de Haas}, Ernst and {Dilday}, Ben and {Doi}, Mamoru and {Eisenstein}, Daniel J. and {Evans}, Michael L. and {Evans}, N.~W. and {Fan}, Xiaohui and {Friedman}, Scott D. and {Frieman}, Joshua A. and {Fukugita}, Masataka and {G{\"a}nsicke}, Boris T. and {Gates}, Evalyn and {Gillespie}, Bruce and {Gilmore}, G. and {Gonzalez}, Belinda and {Gonzalez}, Carlos F. and {Grebel}, Eva K. and {Gunn}, James E. and {Gy{\"o}ry}, Zsuzsanna and {Hall}, Patrick B. and {Harding}, Paul and {Harris}, Frederick H. and {Harvanek}, Michael and {Hawley}, Suzanne L. and {Hayes}, Jeffrey J.~E. and {Heckman}, Timothy M. and {Hendry}, John S. and {Hennessy}, Gregory S. and {Hindsley}, Robert B. and {Hoblitt}, J. and {Hogan}, Craig J. and {Hogg}, David W. and {Holtzman}, Jon A. and {Hyde}, Joseph B. and {Ichikawa}, Shin-ichi and {Ichikawa}, Takashi and {Im}, Myungshin and {Ivezi{\'c}}, {\v{Z}}eljko and {Jester}, Sebastian and {Jiang}, Linhua and {Johnson}, Jennifer A. and {Jorgensen}, Anders M. and {Juri{\'c}}, Mario and {Kent}, Stephen M. and {Kessler}, R. and {Kleinman}, S.~J. and {Knapp}, G.~R. and {Konishi}, Kohki and {Kron}, Richard G. and {Krzesinski}, Jurek and {Kuropatkin}, Nikolay and {Lampeitl}, Hubert and {Lebedeva}, Svetlana and {Lee}, Myung Gyoon and {Lee}, Young Sun and {French Leger}, R. and {L{\'e}pine}, S{\'e}bastien and {Li}, Nolan and {Lima}, Marcos and {Lin}, Huan and {Long}, Daniel C. and {Loomis}, Craig P. and {Loveday}, Jon and {Lupton}, Robert H. and {Magnier}, Eugene and {Malanushenko}, Olena and {Malanushenko}, Viktor and {Mandelbaum}, Rachel and {Margon}, Bruce and {Marriner}, John P. and {Mart{\'\i}nez-Delgado}, David and {Matsubara}, Takahiko and {McGehee}, Peregrine M. and {McKay}, Timothy A. and {Meiksin}, Avery and {Morrison}, Heather L. and {Mullally}, Fergal and {Munn}, Jeffrey A. and {Murphy}, Tara and {Nash}, Thomas and {Nebot}, Ada and {Neilsen}, Jr., Eric H. and {Newberg}, Heidi Jo and {Newman}, Peter R. and {Nichol}, Robert C. and {Nicinski}, Tom and {Nieto-Santisteban}, Maria and {Nitta}, Atsuko and {Okamura}, Sadanori and {Oravetz}, Daniel J. and {Ostriker}, Jeremiah P. and {Owen}, Russell and {Padmanabhan}, Nikhil and {Pan}, Kaike and {Park}, Changbom and {Pauls}, George and {Peoples}, Jr., John and {Percival}, Will J. and {Pier}, Jeffrey R. and {Pope}, Adrian C. and {Pourbaix}, Dimitri and {Price}, Paul A. and {Purger}, Norbert and {Quinn}, Thomas and {Raddick}, M. Jordan and {Re Fiorentin}, Paola and {Richards}, Gordon T. and {Richmond}, Michael W. and {Riess}, Adam G. and {Rix}, Hans-Walter and {Rockosi}, Constance M. and {Sako}, Masao and {Schlegel}, David J. and {Schneider}, Donald P. and {Scholz}, Ralf-Dieter and {Schreiber}, Matthias R. and {Schwope}, Axel D. and {Seljak}, Uro{\v{s}} and {Sesar}, Branimir and {Sheldon}, Erin and {Shimasaku}, Kazu and {Sibley}, Valena C. and {Simmons}, A.~E. and {Sivarani}, Thirupathi and {Allyn Smith}, J. and {Smith}, Martin C. and {Smol{\v{c}}i{\'c}}, Vernesa and {Snedden}, Stephanie A. and {Stebbins}, Albert and {Steinmetz}, Matthias and {Stoughton}, Chris and {Strauss}, Michael A. and {SubbaRao}, Mark and {Suto}, Yasushi and {Szalay}, Alexander S. and {Szapudi}, Istv{\'a}n and {Szkody}, Paula and {Tanaka}, Masayuki and {Tegmark}, Max and {Teodoro}, Luis F.~A. and {Thakar}, Aniruddha R. and {Tremonti}, Christy A. and {Tucker}, Douglas L. and {Uomoto}, Alan and {Vanden Berk}, Daniel E. and {Vandenberg}, Jan and {Vidrih}, S. and {Vogeley}, Michael S. and {Voges}, Wolfgang and {Vogt}, Nicole P. and {Wadadekar}, Yogesh and {Watters}, Shannon and {Weinberg}, David H. and {West}, Andrew A. and {White}, Simon D.~M. and {Wilhite}, Brian C. and {Wonders}, Alainna C. and {Yanny}, Brian and {Yocum}, D.~R.},
	doi = {10.1088/0067-0049/182/2/543},
	eprint = {0812.0649},
	journal = {\apjs},
	month = jun,
	number = {2},
	pages = {543-558},
	primaryclass = {astro-ph},
	title = {{The Seventh Data Release of the Sloan Digital Sky Survey}},
	volume = {182},
	year = 2009}

@article{Corral2011,
	adsurl = {https://ui.adsabs.harvard.edu/abs/2011A&A...530A..42C},
	archiveprefix = {arXiv},
	author = {{Corral}, A. and {Della Ceca}, R. and {Caccianiga}, A. and {Severgnini}, P. and {Brunner}, H. and {Carrera}, F.~J. and {Page}, M.~J. and {Schwope}, A.~D.},
	doi = {10.1051/0004-6361/201015227},
	eid = {A42},
	eprint = {1104.2173},
	journal = {\aap},
	month = jun,
	pages = {A42},
	primaryclass = {astro-ph.CO},
	title = {{The X-ray spectral properties of the AGN population in the XMM-Newton bright serendipitous survey}},
	volume = {530},
	year = 2011}

@article{Malizia2010,
	adsurl = {https://ui.adsabs.harvard.edu/abs/2010MNRAS.408..975M},
	archiveprefix = {arXiv},
	author = {{Malizia}, A. and {Bassani}, L. and {Sguera}, V. and {Stephen}, J.~B. and {Bazzano}, A. and {Fiocchi}, M. and {Bird}, A.~J.},
	doi = {10.1111/j.1365-2966.2010.17157.x},
	eprint = {1006.1272},
	journal = {\mnras},
	month = oct,
	number = {2},
	pages = {975-982},
	primaryclass = {astro-ph.HE},
	title = {{XMM-Newton observations of unidentified INTEGRAL/IBIS sources}},
	volume = {408},
	year = 2010}

@article{Barnard2008,
	adsurl = {https://ui.adsabs.harvard.edu/abs/2008MNRAS.388..849B},
	archiveprefix = {arXiv},
	author = {{Barnard}, R. and {Greening}, L. Shaw and {Kolb}, U.},
	doi = {10.1111/j.1365-2966.2008.13447.x},
	eprint = {0805.1290},
	journal = {\mnras},
	month = aug,
	number = {2},
	pages = {849-862},
	primaryclass = {astro-ph},
	title = {{A multi-coloured survey of NGC 253 with XMM-Newton: testing the methods used for creating luminosity functions from low-count data}},
	volume = {388},
	year = 2008}

@article{Civano2016,
	adsurl = {https://ui.adsabs.harvard.edu/abs/2016ApJ...819...62C},
	archiveprefix = {arXiv},
	author = {{Civano}, F. and {Marchesi}, S. and {Comastri}, A. and {Urry}, M.~C. and {Elvis}, M. and {Cappelluti}, N. and {Puccetti}, S. and {Brusa}, M. and {Zamorani}, G. and {Hasinger}, G. and {Aldcroft}, T. and {Alexander}, D.~M. and {Allevato}, V. and {Brunner}, H. and {Capak}, P. and {Finoguenov}, A. and {Fiore}, F. and {Fruscione}, A. and {Gilli}, R. and {Glotfelty}, K. and {Griffiths}, R.~E. and {Hao}, H. and {Harrison}, F.~A. and {Jahnke}, K. and {Kartaltepe}, J. and {Karim}, A. and {LaMassa}, S.~M. and {Lanzuisi}, G. and {Miyaji}, T. and {Ranalli}, P. and {Salvato}, M. and {Sargent}, M. and {Scoville}, N.~J. and {Schawinski}, K. and {Schinnerer}, E. and {Silverman}, J. and {Smolcic}, V. and {Stern}, D. and {Toft}, S. and {Trakhtenbrot}, B. and {Treister}, E. and {Vignali}, C.},
	doi = {10.3847/0004-637X/819/1/62},
	eid = {62},
	eprint = {1601.00941},
	journal = {\apj},
	month = mar,
	number = {1},
	pages = {62},
	primaryclass = {astro-ph.GA},
	title = {{The Chandra Cosmos Legacy Survey: Overview and Point Source Catalog}},
	volume = {819},
	year = 2016}

@article{Evans2024,
	adsurl = {https://ui.adsabs.harvard.edu/abs/2024ApJS..274...22E},
	archiveprefix = {arXiv},
	author = {{Evans}, Ian N. and {Evans}, Janet D. and {Mart{\'\i}nez-Galarza}, J. Rafael and {Miller}, Joseph B. and {Primini}, Francis A. and {Azadi}, Mojegan and {Burke}, Douglas J. and {Civano}, Francesca M. and {D'Abrusco}, Raffaele and {Fabbiano}, Giuseppina and {Graessle}, Dale E. and {Grier}, John D. and {Houck}, John C. and {Lauer}, Jennifer and {McCollough}, Michael L. and {Nowak}, Michael A. and {Plummer}, David A. and {Rots}, Arnold H. and {Siemiginowska}, Aneta and {Tibbetts}, Michael S.},
	doi = {10.3847/1538-4365/ad6319},
	eid = {22},
	eprint = {2407.10799},
	journal = {\apjs},
	month = oct,
	number = {2},
	pages = {22},
	primaryclass = {astro-ph.HE},
	title = {{The Chandra Source Catalog Release 2 Series}},
	volume = {274},
	year = 2024}

@article{Pounds2006,
	adsurl = {https://ui.adsabs.harvard.edu/abs/2006MNRAS.368..707P},
	archiveprefix = {arXiv},
	author = {{Pounds}, Ken and {Vaughan}, Simon},
	doi = {10.1111/j.1365-2966.2006.10139.x},
	eprint = {astro-ph/0602185},
	journal = {MNRAS},
	month = may,
	number = {2},
	pages = {707-714},
	primaryclass = {astro-ph},
	title = {{X-ray reflection in the nearby Seyfert 2 galaxy NGC 1068}},
	volume = {368},
	year = 2006}

@article{Shen2013,
	adsurl = {https://ui.adsabs.harvard.edu/abs/2013BASI...41...61S},
	archiveprefix = {arXiv},
	author = {{Shen}, Yue},
	doi = {10.48550/arXiv.1302.2643},
	eprint = {1302.2643},
	journal = {Bulletin of the Astronomical Society of India},
	month = mar,
	number = {1},
	pages = {61-115},
	primaryclass = {astro-ph.CO},
	title = {{The mass of quasars}},
	volume = {41},
	year = 2013}

@article{Dexter2019,
	adsurl = {https://ui.adsabs.harvard.edu/abs/2019MNRAS.483L..17D},
	archiveprefix = {arXiv},
	author = {{Dexter}, Jason and {Begelman}, Mitchell C.},
	doi = {10.1093/mnrasl/sly213},
	eprint = {1807.03314},
	journal = {MNRAS},
	month = feb,
	number = {1},
	pages = {L17-L21},
	primaryclass = {astro-ph.GA},
	title = {{Extreme AGN variability: evidence of magnetically elevated accretion?}},
	volume = {483},
	year = 2019}

@article{Comisso2019,
	author = {{Comisso}, Luca and {Sironi}, Lorenzo},
	eid = {122},
	journal = {ApJ},
	month = dec,
	number = {2},
	pages = {122},
	title = {{The Interplay of Magnetically Dominated Turbulence and Magnetic Reconnection in Producing Nonthermal Particles}},
	volume = {886},
	year = 2019}

@article{Dauser2010,
	adsurl = {https://ui.adsabs.harvard.edu/abs/2010MNRAS.409.1534D},
	archiveprefix = {arXiv},
	author = {{Dauser}, T. and {Wilms}, J. and {Reynolds}, C.~S. and {Brenneman}, L.~W.},
	doi = {10.1111/j.1365-2966.2010.17393.x},
	eprint = {1007.4937},
	journal = {MNRAS},
	month = dec,
	number = {4},
	pages = {1534-1540},
	primaryclass = {astro-ph.HE},
	title = {{Broad emission lines for a negatively spinning black hole}},
	volume = {409},
	year = 2010}

@article{Garcia2010,
	adsurl = {https://ui.adsabs.harvard.edu/abs/2010ApJ...718..695G},
	archiveprefix = {arXiv},
	author = {{Garc{\'\i}a}, J. and {Kallman}, T.~R.},
	doi = {10.1088/0004-637X/718/2/695},
	eprint = {1006.0485},
	journal = {ApJ},
	month = aug,
	number = {2},
	pages = {695-706},
	primaryclass = {astro-ph.HE},
	title = {{X-ray Reflected Spectra from Accretion Disk Models. I. Constant Density Atmospheres}},
	volume = {718},
	year = 2010}

@article{Fabian2000,
	adsurl = {https://ui.adsabs.harvard.edu/abs/2000PASP..112.1145F},
	archiveprefix = {arXiv},
	author = {{Fabian}, A.~C. and {Iwasawa}, K. and {Reynolds}, C.~S. and {Young}, A.~J.},
	doi = {10.1086/316610},
	eprint = {astro-ph/0004366},
	journal = {PASP},
	month = sep,
	number = {775},
	pages = {1145-1161},
	primaryclass = {astro-ph},
	title = {{Broad Iron Lines in Active Galactic Nuclei}},
	volume = {112},
	year = 2000}

@article{Ulrich1997,
	adsurl = {https://ui.adsabs.harvard.edu/abs/1997ARA&A..35..445U},
	author = {{Ulrich}, Marie-Helene and {Maraschi}, Laura and {Urry}, C. Megan},
	doi = {10.1146/annurev.astro.35.1.445},
	journal = {ARA\&A},
	month = jan,
	pages = {445-502},
	title = {{Variability of Active Galactic Nuclei}},
	volume = {35},
	year = 1997}

@article{Mushotzky1993,
	adsurl = {https://ui.adsabs.harvard.edu/abs/1993ARA&A..31..717M},
	author = {{Mushotzky}, Richard F. and {Done}, Christine and {Pounds}, Kenneth A.},
	doi = {10.1146/annurev.aa.31.090193.00344110.1146/annurev.astro.31.1.717},
	journal = {ARA\&A},
	month = jan,
	pages = {717-717},
	title = {{X-ray spectra and time variability of active galactic nuclei.}},
	volume = {31},
	year = 1993}

@article{Elvis1994,
	adsurl = {https://ui.adsabs.harvard.edu/abs/1994ApJS...95....1E},
	author = {{Elvis}, Martin and {Wilkes}, Belinda J. and {McDowell}, Jonathan C. and {Green}, Richard F. and {Bechtold}, Jill and {Willner}, S.~P. and {Oey}, M.~S. and {Polomski}, Elisha and {Cutri}, Roc},
	doi = {10.1086/192093},
	journal = {ApJS},
	month = nov,
	pages = {1},
	title = {{Atlas of Quasar Energy Distributions}},
	volume = {95},
	year = 1994}

@article{Kara2025,
	adsurl = {https://ui.adsabs.harvard.edu/abs/2025ARA&A..63..379K},
	archiveprefix = {arXiv},
	author = {{Kara}, Erin and {Garc{\'\i}a}, Javier},
	doi = {10.1146/annurev-astro-071221-052844},
	eprint = {2503.22791},
	journal = {ARA\&A},
	month = aug,
	number = {1},
	pages = {379-430},
	primaryclass = {astro-ph.HE},
	title = {{Supermassive Black Holes in X-Rays: From Standard Accretion to Extreme Transients}},
	volume = {63},
	year = 2025}

@article{Duras2020,
	adsurl = {https://ui.adsabs.harvard.edu/abs/2020A&A...636A..73D},
	archiveprefix = {arXiv},
	author = {{Duras}, F. and {Bongiorno}, A. and {Ricci}, F. and {Piconcelli}, E. and {Shankar}, F. and {Lusso}, E. and {Bianchi}, S. and {Fiore}, F. and {Maiolino}, R. and {Marconi}, A. and {Onori}, F. and {Sani}, E. and {Schneider}, R. and {Vignali}, C. and {La Franca}, F.},
	doi = {10.1051/0004-6361/201936817},
	eid = {A73},
	eprint = {2001.09984},
	journal = {A\&A},
	month = apr,
	pages = {A73},
	primaryclass = {astro-ph.GA},
	title = {{Universal bolometric corrections for active galactic nuclei over seven luminosity decades}},
	volume = {636},
	year = 2020}

@article{Lusso2012,
	adsurl = {https://ui.adsabs.harvard.edu/abs/2012MNRAS.425..623L},
	archiveprefix = {arXiv},
	author = {{Lusso}, E. and {Comastri}, A. and {Simmons}, B.~D. and {Mignoli}, M. and {Zamorani}, G. and {Vignali}, C. and {Brusa}, M. and {Shankar}, F. and {Lutz}, D. and {Trump}, J.~R. and {Maiolino}, R. and {Gilli}, R. and {Bolzonella}, M. and {Puccetti}, S. and {Salvato}, M. and {Impey}, C.~D. and {Civano}, F. and {Elvis}, M. and {Mainieri}, V. and {Silverman}, J.~D. and {Koekemoer}, A.~M. and {Bongiorno}, A. and {Merloni}, A. and {Berta}, S. and {Le Floc'h}, E. and {Magnelli}, B. and {Pozzi}, F. and {Riguccini}, L.},
	doi = {10.1111/j.1365-2966.2012.21513.x},
	eprint = {1206.2642},
	journal = {MNRAS},
	month = sep,
	number = {1},
	pages = {623-640},
	primaryclass = {astro-ph.CO},
	title = {{Bolometric luminosities and Eddington ratios of X-ray selected active galactic nuclei in the XMM-COSMOS survey}},
	volume = {425},
	year = 2012}

@article{Marchesi2016,
	adsurl = {https://ui.adsabs.harvard.edu/abs/2016ApJ...830..100M},
	archiveprefix = {arXiv},
	author = {{Marchesi}, S. and {Lanzuisi}, G. and {Civano}, F. and {Iwasawa}, K. and {Suh}, H. and {Comastri}, A. and {Zamorani}, G. and {Allevato}, V. and {Griffiths}, R. and {Miyaji}, T. and {Ranalli}, P. and {Salvato}, M. and {Schawinski}, K. and {Silverman}, J. and {Treister}, E. and {Urry}, C.~M. and {Vignali}, C.},
	doi = {10.3847/0004-637X/830/2/100},
	eid = {100},
	eprint = {1608.05149},
	journal = {ApJ},
	month = oct,
	number = {2},
	pages = {100},
	primaryclass = {astro-ph.GA},
	title = {{The Chandra COSMOS-Legacy Survey: Source X-Ray Spectral Properties}},
	volume = {830},
	year = 2016}

@article{Mauduit2012,
	adsurl = {https://ui.adsabs.harvard.edu/abs/2012PASP..124..714M},
	archiveprefix = {arXiv},
	author = {{Mauduit}, J. -C. and {Lacy}, M. and {Farrah}, D. and {Surace}, J.~A. and {Jarvis}, M. and {Oliver}, S. and {Maraston}, C. and {Vaccari}, M. and {Marchetti}, L. and {Zeimann}, G. and {Gonz{\'a}les-Solares}, E.~A. and {Pforr}, J. and {Petric}, A.~O. and {Henriques}, B. and {Thomas}, P.~A. and {Afonso}, J. and {Rettura}, A. and {Wilson}, G. and {Falder}, J.~T. and {Geach}, J.~E. and {Huynh}, M. and {Norris}, R.~P. and {Seymour}, N. and {Richards}, G.~T. and {Stanford}, S.~A. and {Alexander}, D.~M. and {Becker}, R.~H. and {Best}, P.~N. and {Bizzocchi}, L. and {Bonfield}, D. and {Castro}, N. and {Cava}, A. and {Chapman}, S. and {Christopher}, N. and {Clements}, D.~L. and {Covone}, G. and {Dubois}, N. and {Dunlop}, J.~S. and {Dyke}, E. and {Edge}, A. and {Ferguson}, H.~C. and {Foucaud}, S. and {Franceschini}, A. and {Gal}, R.~R. and {Grant}, J.~K. and {Grossi}, M. and {Hatziminaoglou}, E. and {Hickey}, S. and {Hodge}, J.~A. and {Huang}, J. -S. and {Ivison}, R.~J. and {Kim}, M. and {LeFevre}, O. and {Lehnert}, M. and {Lonsdale}, C.~J. and {Lubin}, L.~M. and {McLure}, R.~J. and {Messias}, H. and {Mart{\'\i}nez-Sansigre}, A. and {Mortier}, A.~M.~J. and {Nielsen}, D.~M. and {Ouchi}, M. and {Parish}, G. and {Perez-Fournon}, I. and {Pierre}, M. and {Rawlings}, S. and {Readhead}, A. and {Ridgway}, S.~E. and {Rigopoulou}, D. and {Romer}, A.~K. and {Rosebloom}, I.~G. and {Rottgering}, H.~J.~A. and {Rowan-Robinson}, M. and {Sajina}, A. and {Simpson}, C.~J. and {Smail}, I. and {Squires}, G.~K. and {Stevens}, J.~A. and {Taylor}, R. and {Trichas}, M. and {Urrutia}, T. and {van Kampen}, E. and {Verma}, A. and {Xu}, C.~K.},
	doi = {10.1086/666945},
	eprint = {1206.4060},
	journal = {PASP},
	month = jul,
	number = {917},
	pages = {714},
	primaryclass = {astro-ph.CO},
	title = {{The Spitzer Extragalactic Representative Volume Survey (SERVS): Survey Definition and Goals}},
	volume = {124},
	year = 2012}

@article{Wilkes2009,
	adsurl = {https://ui.adsabs.harvard.edu/abs/2009ApJS..185..433W},
	archiveprefix = {arXiv},
	author = {{Wilkes}, Belinda J. and {Kilgard}, Roy and {Kim}, Dong-Woo and {Kim}, Minsun and {Polletta}, Mari and {Lonsdale}, Carol and {Smith}, Harding E. and {Surace}, Jason and {Owen}, Frazer N. and {Franceschini}, A. and {Siana}, Brian and {Shupe}, David},
	doi = {10.1088/0067-0049/185/2/433},
	eprint = {0910.2675},
	journal = {ApJS},
	month = dec,
	number = {2},
	pages = {433-450},
	primaryclass = {astro-ph.HE},
	title = {{The SWIRE/Chandra Survey: The X-ray Sources}},
	volume = {185},
	year = 2009}

@article{Stern2018,
	adsurl = {https://ui.adsabs.harvard.edu/abs/2018ApJ...864...27S},
	archiveprefix = {arXiv},
	author = {{Stern}, Daniel and {McKernan}, Barry and {Graham}, Matthew J. and {Ford}, K.~E.~S. and {Ross}, Nicholas P. and {Meisner}, Aaron M. and {Assef}, Roberto J. and {Balokovi{\'c}}, Mislav and {Brightman}, Murray and {Dey}, Arjun and {Drake}, Andrew and {Djorgovski}, S.~G. and {Eisenhardt}, Peter and {Jun}, Hyunsung D.},
	doi = {10.3847/1538-4357/aac726},
	eid = {27},
	eprint = {1805.06920},
	journal = {ApJ},
	month = sep,
	number = {1},
	pages = {27},
	primaryclass = {astro-ph.GA},
	title = {{A Mid-IR Selected Changing-look Quasar and Physical Scenarios for Abrupt AGN Fading}},
	volume = {864},
	year = 2018}

@article{Rowan2017,
	author = {{Rowan}, Michael E. and {Sironi}, Lorenzo and {Narayan}, Ramesh},
	eid = {29},
	journal = {ApJ},
	month = nov,
	number = {1},
	pages = {29},
	title = {{Electron and Proton Heating in Transrelativistic Magnetic Reconnection}},
	volume = {850},
	year = 2017}

@article{Ricci2017,
	adsurl = {https://ui.adsabs.harvard.edu/abs/2017ApJS..233...17R},
	archiveprefix = {arXiv},
	author = {{Ricci}, C. and {Trakhtenbrot}, B. and {Koss}, M.~J. and {Ueda}, Y. and {Delvecchio}, I. and {Treister}, E. and {Schawinski}, K. and {Paltani}, S. and {Oh}, K. and {Lamperti}, I. and {Berney}, S. and {Gandhi}, P. and {Ichikawa}, K. and {Bauer}, F.~E. and {Ho}, L.~C. and {Asmus}, D. and {Beckmann}, V. and {Soldi}, S. and {Balokovi{\'c}}, M. and {Gehrels}, N. and {Markwardt}, C.~B.},
	doi = {10.3847/1538-4365/aa96ad},
	eid = {17},
	eprint = {1709.03989},
	journal = {ApJS},
	month = dec,
	number = {2},
	pages = {17},
	primaryclass = {astro-ph.HE},
	title = {{BAT AGN Spectroscopic Survey. V. X-Ray Properties of the Swift/BAT 70-month AGN Catalog}},
	volume = {233},
	year = 2017}

@article{Yuan2025,
	adsurl = {https://ui.adsabs.harvard.edu/abs/2025SCPMA..6839501Y},
	archiveprefix = {arXiv},
	author = {{Yuan}, Weimin and {Dai}, Lixin and {Feng}, Hua and {Jin}, Chichuan and {Jonker}, Peter and {Kuulkers}, Erik and {Liu}, Yuan and {Nandra}, Kirpal and {O'Brien}, Paul and {Piro}, Luigi and {Rau}, Arne and {Rea}, Nanda and {Sanders}, Jeremy and {Tao}, Lian and {Wang}, Junfeng and {Wu}, Xuefeng and {Zhang}, Bing and {Zhang}, Shuangnan and {Ai}, Shunke and {Buchner}, Johannes and {Bulbul}, Esra and {Chen}, Hechao and {Chen}, Minghua and {Chen}, Yong and {Chen}, Yu-Peng and {Coleiro}, Alexis and {Zelati}, Francesco Coti and {Dai}, Zigao and {Fan}, Xilong and {Fan}, Zhou and {Friedrich}, Susanne and {Gao}, He and {Ge}, Chong and {Ge}, Mingyu and {Geng}, Jinjun and {Ghirlanda}, Giancarlo and {Gianfagna}, Giulia and {Gou}, Lijun and {Guillot}, S{\'e}bastien and {Hou}, Xian and {Hu}, Jingwei and {Huang}, Yongfeng and {Ji}, Long and {Jia}, Shumei and {Komossa}, S. and {Kong}, Albert K.~H. and {Lan}, Lin and {Li}, An and {Li}, Ang and {Li}, Chengkui and {Li}, Dongyue and {Li}, Jian and {Li}, Zhaosheng and {Ling}, Zhixing and {Liu}, Ang and {Liu}, Jinzhong and {Liu}, Liangduan and {Liu}, Zhu and {Luo}, Jiawei and {Ma}, Ruican and {Maggi}, Pierre and {Maitra}, Chandreyee and {Marino}, Alessio and {Ng}, Stephen Chi-Yung and {Pan}, Haiwu and {Rukdee}, Surangkhana and {Soria}, Roberto and {Sun}, Hui and {Tam}, Pak-Hin Thomas and {Thakur}, Aishwarya Linesh and {Tian}, Hui and {Troja}, Eleonora and {Wang}, Wei and {Wang}, Xiangyu and {Wang}, Yanan and {Wei}, Junjie and {Wen}, Sixiang and {Wu}, Jianfeng and {Wu}, Ting and {Xiao}, Di and {Xu}, Dong and {Xu}, Renxin and {Xu}, Yanjun and {Xu}, Yu and {Yang}, Haonan and {You}, Bei and {Yu}, Heng and {Yu}, Yunwei and {Zhang}, Binbin and {Zhang}, Chen and {Zhang}, Guobao and {Zhang}, Liang and {Zhang}, Wenda and {Zhang}, Yu and {Zhou}, Ping and {Zou}, Zecheng},
	doi = {10.1007/s11433-024-2600-3},
	eid = {239501},
	eprint = {2501.07362},
	journal = {Science China Physics, Mechanics, and Astronomy},
	month = mar,
	number = {3},
	pages = {239501},
	primaryclass = {astro-ph.HE},
	title = {{Science objectives of the Einstein Probe mission}},
	volume = {68},
	year = 2025}

@article{Pandey2023,
	adsurl = {https://ui.adsabs.harvard.edu/abs/2023A&A...680A.102P},
	archiveprefix = {arXiv},
	author = {{Pandey}, Ashwani and {Czerny}, Bo{\.z}ena and {Panda}, Swayamtrupta and {Prince}, Raj and {Jaiswal}, Vikram Kumar and {Martinez-Aldama}, Mary Loli and {Zaja{\v{c}}ek}, Michal and {{\'S}niegowska}, Marzena},
	doi = {10.1051/0004-6361/202347819},
	eid = {A102},
	eprint = {2310.05089},
	journal = {A\&A},
	month = dec,
	pages = {A102},
	primaryclass = {astro-ph.GA},
	title = {{Broad-line region in active galactic nuclei: Dusty or dustless?}},
	volume = {680},
	year = 2023}

@article{Maiolino2001,
	adsurl = {https://ui.adsabs.harvard.edu/abs/2001A&A...375...25M},
	archiveprefix = {arXiv},
	author = {{Maiolino}, R. and {Salvati}, M. and {Marconi}, A. and {Antonucci}, R.~R.~J.},
	doi = {10.1051/0004-6361:20010808},
	eprint = {astro-ph/0106192},
	journal = {A\&A},
	month = aug,
	pages = {25-29},
	primaryclass = {astro-ph},
	title = {{The Ly-edge paradox and the need for obscured QSOs}},
	volume = {375},
	year = 2001}

@article{Garcia2014,
	adsurl = {https://ui.adsabs.harvard.edu/abs/2014ApJ...782...76G},
	archiveprefix = {arXiv},
	author = {{Garc{\'\i}a}, J. and {Dauser}, T. and {Lohfink}, A. and {Kallman}, T.~R. and {Steiner}, J.~F. and {McClintock}, J.~E. and {Brenneman}, L. and {Wilms}, J. and {Eikmann}, W. and {Reynolds}, C.~S. and {Tombesi}, F.},
	doi = {10.1088/0004-637X/782/2/76},
	eid = {76},
	eprint = {1312.3231},
	journal = {ApJ},
	month = feb,
	number = {2},
	pages = {76},
	primaryclass = {astro-ph.HE},
	title = {{Improved Reflection Models of Black Hole Accretion Disks: Treating the Angular Distribution of X-Rays}},
	volume = {782},
	year = 2014}

@article{Uttley2014,
	adsurl = {https://ui.adsabs.harvard.edu/abs/2014A&ARv..22...72U},
	archiveprefix = {arXiv},
	author = {{Uttley}, P. and {Cackett}, E.~M. and {Fabian}, A.~C. and {Kara}, E. and {Wilkins}, D.~R.},
	doi = {10.1007/s00159-014-0072-0},
	eid = {72},
	eprint = {1405.6575},
	journal = {A\&ARv},
	month = aug,
	pages = {72},
	primaryclass = {astro-ph.HE},
	title = {{X-ray reverberation around accreting black holes}},
	volume = {22},
	year = 2014}

@article{Dai2010,
	adsurl = {https://ui.adsabs.harvard.edu/abs/2010ApJ...709..278D},
	archiveprefix = {arXiv},
	author = {{Dai}, X. and {Kochanek}, C.~S. and {Chartas}, G. and {Koz{\l}owski}, S. and {Morgan}, C.~W. and {Garmire}, G. and {Agol}, E.},
	doi = {10.1088/0004-637X/709/1/278},
	eprint = {0906.4342},
	journal = {ApJ},
	month = jan,
	number = {1},
	pages = {278-285},
	primaryclass = {astro-ph.HE},
	title = {{The Sizes of the X-ray and Optical Emission Regions of RXJ 1131-1231}},
	volume = {709},
	year = 2010}

@article{Chartas2009,
	adsurl = {https://ui.adsabs.harvard.edu/abs/2009ApJ...693..174C},
	archiveprefix = {arXiv},
	author = {{Chartas}, G. and {Kochanek}, C.~S. and {Dai}, X. and {Poindexter}, S. and {Garmire}, G.},
	doi = {10.1088/0004-637X/693/1/174},
	eprint = {0805.4492},
	journal = {ApJ},
	month = mar,
	number = {1},
	pages = {174-185},
	primaryclass = {astro-ph},
	title = {{X-Ray Microlensing in RXJ1131-1231 and HE1104-1805}},
	volume = {693},
	year = 2009}

@article{Levenson2001,
	adsurl = {https://ui.adsabs.harvard.edu/abs/2001ApJ...550..230L},
	archiveprefix = {arXiv},
	author = {{Levenson}, N.~A. and {Weaver}, K.~A. and {Heckman}, T.~M.},
	doi = {10.1086/319726},
	eprint = {astro-ph/0012036},
	journal = {ApJ},
	month = mar,
	number = {1},
	pages = {230-242},
	primaryclass = {astro-ph},
	title = {{The Seyfert-Starburst Connection in X-Rays. II. Results and Implications}},
	volume = {550},
	year = 2001}

@article{Magdziarz1998,
	adsurl = {https://ui.adsabs.harvard.edu/abs/1998MNRAS.301..179M},
	author = {{Magdziarz}, Pawel and {Blaes}, Omer M. and {Zdziarski}, Andrzej A. and {Johnson}, W. Neil and {Smith}, David A.},
	doi = {10.1046/j.1365-8711.1998.02015.x},
	journal = {MNRAS},
	month = nov,
	number = {1},
	pages = {179-192},
	title = {{A spectral decomposition of the variable optical, ultraviolet and X-ray continuum of NGC 5548}},
	volume = {301},
	year = 1998}

@article{Ballantyne2001,
	adsurl = {https://ui.adsabs.harvard.edu/abs/2001MNRAS.323..506B},
	archiveprefix = {arXiv},
	author = {{Ballantyne}, D.~R. and {Iwasawa}, K. and {Fabian}, A.~C.},
	doi = {10.1046/j.1365-8711.2001.04234.x},
	eprint = {astro-ph/0011360},
	journal = {MNRAS},
	month = may,
	number = {2},
	pages = {506-516},
	primaryclass = {astro-ph},
	title = {{Evidence for ionized accretion discs in five narrow-line Seyfert 1 galaxies}},
	volume = {323},
	year = 2001}

@article{Turner1989,
	adsurl = {https://ui.adsabs.harvard.edu/abs/1989MNRAS.240..833T},
	author = {{Turner}, T.~J. and {Pounds}, K.~A.},
	doi = {10.1093/mnras/240.4.833},
	journal = {MNRAS},
	month = oct,
	pages = {833-880},
	title = {{The EXOSAT spectral survey of AGN.}},
	volume = {240},
	year = 1989}

@article{Jiang2018,
	adsurl = {https://ui.adsabs.harvard.edu/abs/2018MNRAS.477.3711J},
	archiveprefix = {arXiv},
	author = {{Jiang}, J. and {Parker}, M.~L. and {Fabian}, A.~C. and {Alston}, W.~N. and {Buisson}, D.~J.~K. and {Cackett}, E.~M. and {Chiang}, C. -Y. and {Dauser}, T. and {Gallo}, L.~C. and {Garc{\'\i}a}, J.~A. and {Harrison}, F.~A. and {Lohfink}, A.~M. and {De Marco}, B. and {Kara}, E. and {Miller}, J.~M. and {Miniutti}, G. and {Pinto}, C. and {Walton}, D.~J. and {Wilkins}, D.~R.},
	doi = {10.1093/mnras/sty836},
	eprint = {1804.00349},
	journal = {MNRAS},
	month = jul,
	number = {3},
	pages = {3711-3726},
	primaryclass = {astro-ph.HE},
	title = {{The 1.5 Ms observing campaign on IRAS 13224-3809 - I. X-ray spectral analysis}},
	volume = {477},
	year = 2018}

@article{Risaliti2013,
	adsurl = {https://ui.adsabs.harvard.edu/abs/2013Natur.494..449R},
	archiveprefix = {arXiv},
	author = {{Risaliti}, G. and {Harrison}, F.~A. and {Madsen}, K.~K. and {Walton}, D.~J. and {Boggs}, S.~E. and {Christensen}, F.~E. and {Craig}, W.~W. and {Grefenstette}, B.~W. and {Hailey}, C.~J. and {Nardini}, E. and {Stern}, Daniel and {Zhang}, W.~W.},
	doi = {10.1038/nature11938},
	eprint = {1302.7002},
	journal = {Nature},
	month = feb,
	number = {7438},
	pages = {449-451},
	primaryclass = {astro-ph.HE},
	title = {{A rapidly spinning supermassive black hole at the centre of NGC 1365}},
	volume = {494},
	year = 2013}

@article{Walton2021,
	adsurl = {https://ui.adsabs.harvard.edu/abs/2021MNRAS.506.1557W},
	archiveprefix = {arXiv},
	author = {{Walton}, D.~J. and {Balokovi{\'c}}, M. and {Fabian}, A.~C. and {Gallo}, L.~C. and {Koss}, M. and {Nardini}, E. and {Reynolds}, C.~S. and {Ricci}, C. and {Stern}, D. and {Alston}, W.~N. and {Dauser}, T. and {Garc{\'\i}a}, J.~A. and {Kosec}, P. and {Reynolds}, M.~T. and {Harrison}, F.~A. and {Miller}, J.~M.},
	doi = {10.1093/mnras/stab1290},
	eprint = {2107.10278},
	journal = {MNRAS},
	month = sep,
	number = {2},
	pages = {1557-1572},
	primaryclass = {astro-ph.HE},
	title = {{Extreme relativistic reflection in the active galaxy ESO 033-G002}},
	volume = {506},
	year = 2021}

@article{Reynolds2003,
	adsurl = {https://ui.adsabs.harvard.edu/abs/2003PhR...377..389R},
	archiveprefix = {arXiv},
	author = {{Reynolds}, Christopher S. and {Nowak}, Michael A.},
	doi = {10.1016/S0370-1573(02)00584-7},
	eprint = {astro-ph/0212065},
	journal = {Physics Reports},
	month = apr,
	number = {6},
	pages = {389-466},
	primaryclass = {astro-ph},
	title = {{Fluorescent iron lines as a probe of astrophysical black hole systems}},
	volume = {377},
	year = 2003}

@article{Fabian2009,
	adsurl = {https://ui.adsabs.harvard.edu/abs/2009Natur.459..540F},
	author = {{Fabian}, A.~C. and {Zoghbi}, A. and {Ross}, R.~R. and {Uttley}, P. and {Gallo}, L.~C. and {Brandt}, W.~N. and {Blustin}, A.~J. and {Boller}, T. and {Caballero-Garcia}, M.~D. and {Larsson}, J. and {Miller}, J.~M. and {Miniutti}, G. and {Ponti}, G. and {Reis}, R.~C. and {Reynolds}, C.~S. and {Tanaka}, Y. and {Young}, A.~J.},
	doi = {10.1038/nature08007},
	journal = {Nature},
	month = may,
	number = {7246},
	pages = {540-542},
	title = {{Broad line emission from iron K- and L-shell transitions in the active galaxy 1H0707-495}},
	volume = {459},
	year = 2009}

@article{Tanaka1995,
	adsurl = {https://ui.adsabs.harvard.edu/abs/1995Natur.375..659T},
	author = {{Tanaka}, Y. and {Nandra}, K. and {Fabian}, A.~C. and {Inoue}, H. and {Otani}, C. and {Dotani}, T. and {Hayashida}, K. and {Iwasawa}, K. and {Kii}, T. and {Kunieda}, H. and {Makino}, F. and {Matsuoka}, M.},
	doi = {10.1038/375659a0},
	journal = {Nature},
	month = jun,
	number = {6533},
	pages = {659-661},
	title = {{Gravitationally redshifted emission implying an accretion disk and massive black hole in the active galaxy MCG-6-30-15}},
	volume = {375},
	year = 1995}

@article{Nandra2007,
	adsurl = {https://ui.adsabs.harvard.edu/abs/2007MNRAS.382..194N},
	archiveprefix = {arXiv},
	author = {{Nandra}, K. and {O'Neill}, P.~M. and {George}, I.~M. and {Reeves}, J.~N.},
	doi = {10.1111/j.1365-2966.2007.12331.x},
	eprint = {0708.1305},
	journal = {MNRAS},
	month = nov,
	number = {1},
	pages = {194-228},
	primaryclass = {astro-ph},
	title = {{An XMM-Newton survey of broad iron lines in Seyfert galaxies}},
	volume = {382},
	year = 2007}

@article{Shen2011,
	adsurl = {https://ui.adsabs.harvard.edu/abs/2011ApJS..194...45S},
	archiveprefix = {arXiv},
	author = {{Shen}, Yue and {Richards}, Gordon T. and {Strauss}, Michael A. and {Hall}, Patrick B. and {Schneider}, Donald P. and {Snedden}, Stephanie and {Bizyaev}, Dmitry and {Brewington}, Howard and {Malanushenko}, Viktor and {Malanushenko}, Elena and {Oravetz}, Dan and {Pan}, Kaike and {Simmons}, Audrey},
	doi = {10.1088/0067-0049/194/2/45},
	eid = {45},
	eprint = {1006.5178},
	journal = {ApJS},
	month = jun,
	number = {2},
	pages = {45},
	primaryclass = {astro-ph.CO},
	title = {{A Catalog of Quasar Properties from Sloan Digital Sky Survey Data Release 7}},
	volume = {194},
	year = 2011}

@article{Zhang2025,
	adsurl = {https://ui.adsabs.harvard.edu/abs/2025arXiv250608101Z},
	archiveprefix = {arXiv},
	author = {{Zhang}, Shuang-Nan and {Santangelo}, Andrea and {Xu}, Yupeng and {Feng}, Hua and {Lu}, Fangjun and {Chen}, Yong and {Ge}, Mingyu and {Nandra}, Kirpal and {Wu}, Xin and {Feroci}, Marco and {Hernanz}, Margarita and {Liu}, Congzhan and {He}, Huilin and {Wang}, Yusa and {Jiang}, Weichun and {Cui}, Weiwei and {Yang}, Yanji and {Wang}, Juan and {Li}, Wei and {Liu}, Xiaohua and {Meng}, Bin and {Wen}, Xiangyang and {Zhang}, Aimei and {Ma}, Jia and {Li}, Maoshun and {Li}, Gang and {Qi}, Liqiang and {Sun}, Jianchao and {Luo}, Tao and {Liu}, Hongwei and {Liu}, Xiaojing and {Zhang}, Fan and {Luo}, Laidan and {Zhu}, Yuxuan and {Zhao}, Zijian and {Sun}, Liang and {Yang}, Xiongtao and {Wu}, Qiong and {Jiang}, Jiechen and {Shi}, Haoli and {Liu}, Jiangtao and {Xu}, Yanbing and {Yang}, Sheng and {Zhang}, Laiyu and {Han}, Dawei and {Gao}, Na and {Huo}, Jia and {Zhang}, Ziliang and {Wang}, Hao and {Zhao}, Xiaofan and {Cui}, Weiwei and {Wang}, Juan and {Wang}, Shuo and {Li}, Zhenjie and {Bao}, Ziyu and {Liu}, Yaoguang and {Wang}, Ke and {Wang}, Na and {Wang}, Bo and {Wang}, Langping and {Wang}, Dianlong and {Ding}, Fei and {Sheng}, Lizhi and {Qiang}, Pengfei and {Yan}, Yongqing and {Liu}, Yongan and {Wu}, Zhenyu and {Liu}, Yichen and {Chen}, Hao and {Zhang}, Yacong and {Liu}, Hongbang and {Altmann}, Alexander and {Bechteler}, Thomas and {Burwitz}, Vadim and {Fiorini}, Carlo and {Friedrich}, Peter and {Meidinger}, Norbert and {Strecker}, Rafael and {Baldini}, Luca and {Bellazzini}, Ronaldo and {Bonino}, Raffaella and {Frass{\`a}}, Andrea and {Latronico}, Luca and {Maldera}, Simone and {Manfreda}, Alberto and {Minuti}, Massimo and {Pesce-Rollins}, Melissa and {Sgr{\`o}}, Carmelo and {Tugliani}, Stefano and {Pareschi}, Giovanni and {Basso}, Stefano and {Sironi}, Giorgia and {Spiga}, Daniele and {Tagliaferri}, Gianpiero and {Tykhonov}, Andrii and {Paltani}, St{\`e}phane and {Bozzo}, Enrico and {Tenzer}, Christoph and {Bayer}, J{\"o}rg and {Tuo}, Youli and {Liu}, Honghui and {Zhang}, Yonghe and {Cai}, Zhiming and {Liu}, Huaqiu and {Chen}, Wen and {Wang}, Chunhong and {He}, Tao and {Chen}, Yehai and {Qiu}, Chengbo and {Zhang}, Ye and {Feng}, Jianchao and {Zhu}, Xiaofei and {Zhou}, Heng and {Zheng}, Shijie and {Song}, Liming and {Shi}, Haoli and {Wang}, Jinzhou and {Jia}, Shumei and {Jiang}, Zewen and {Li}, Xiaobo and {Zhao}, Haisheng and {Guan}, Ju and {Zhang}, Juan and {Li}, Chengkui and {Huang}, Yue and {Liao}, Jinyuan and {You}, Yuan and {Zhang}, Hongmei and {Wang}, Wenshuai and {Wang}, Shuang and {Ou}, Ge and {Hu}, Hao and {Shi}, Jingyan and {Cui}, Tao and {Jiang}, Xiaowei and {Cheng}, Yaodong and {Li}, Haibo and {Xu}, Yanjun and {Zane}, Silvia and {Bambi}, Cosimo and {Bu}, Qingcui and {Dall'Osso}, Simone and {De Rosa}, Alessandra and {Gou}, Lijun and {Guillot}, Sebastien and {Ji}, Long and {Li}, Ang and {Mao}, Jirong and {Patruno}, Alessandro and {Stratta}, Giulia and {Taverna}, Roberto and {Tsygankov}, Sergey and {Uttley}, Phil and {Watts}, Anna L. and {Wu}, Xuefeng and {Xu}, Renxin and {Yi}, Shuxu and {Zhang}, Guobao and {Zhang}, Liang and {Zhao}, Wen and {Zhou}, Ping},
	doi = {10.48550/arXiv.2506.08101},
	eid = {arXiv:2506.08101},
	eprint = {2506.08101},
	journal = {arXiv e-prints},
	month = jun,
	pages = {arXiv:2506.08101},
	primaryclass = {astro-ph.HE},
	title = {{The enhanced X-ray Timing and Polarimetry mission -- eXTP for launch in 2030}},
	year = 2025}

@inproceedings{Reynolds2023,
	adsurl = {https://ui.adsabs.harvard.edu/abs/2023SPIE12678E..1ER},
	archiveprefix = {arXiv},
	author = {{Reynolds}, Christopher S. and {Kara}, Erin A. and {Mushotzky}, Richard F. and {Ptak}, Andrew and {Koss}, Michael J. and {Williams}, Brian J. and {Allen}, Steven W. and {Bauer}, Franz E. and {Bautz}, Marshall and {Bogadhee}, Arash and {Burdge}, Kevin B. and {Cappelluti}, Nico and {Cenko}, Brad and {Chartas}, George and {Chan}, Kai-Wing and {Corrales}, L{\'\i}a. and {Daylan}, Tansu and {Falcone}, Abraham D. and {Foord}, Adi and {Grant}, Catherine E. and {Habouzit}, M{\'e}lanie and {Haggard}, Daryl and {Herrmann}, Sven and {Hodges-Kluck}, Edmund and {Kargaltsev}, Oleg and {King}, George W. and {Kounkel}, Marina and {Lopez}, Laura A. and {Marchesi}, Stefano and {McDonald}, Michael and {Meyer}, Eileen and {Miller}, Eric D. and {Nynka}, Melania and {Okajima}, Takashi and {Pacucci}, Fabio and {Russell}, Helen R. and {Safi-Harb}, Samar and {Strassun}, Keivan G. and {Trindade Falc{\~a}o}, Anna and {Walker}, Stephen A. and {Wilms}, Joern and {Yukita}, Mihoko and {Zhang}, William W.},
	booktitle = {UV, X-Ray, and Gamma-Ray Space Instrumentation for Astronomy XXIII},
	doi = {10.1117/12.2677468},
	editor = {{Siegmund}, Oswald H. and {Hoadley}, Keri},
	eid = {126781E},
	eprint = {2311.00780},
	month = oct,
	pages = {126781E},
	primaryclass = {astro-ph.IM},
	series = {Society of Photo-Optical Instrumentation Engineers (SPIE) Conference Series},
	title = {{Overview of the advanced x-ray imaging satellite (AXIS)}},
	volume = {12678},
	year = 2023}

@article{Marchesi2018,
	author = {S. Marchesi and M. Ajello and L. Marcotulli and A. Comastri and G. Lanzuisi and C. Vignali},
	journal = {ApJ},
	number = {1},
	pages = {49},
	title = {Compton-thick AGNs in the NuSTAR Era},
	url = {http://stacks.iop.org/0004-637X/854/i=1/a=49},
	volume = {854},
	year = {2018}}
\bibliographystyle{aasjournal}

\end{document}